\documentclass[aps,prl,10pt,twocolumn,superscriptaddress,english,longbibliography]{revtex4-2}
\usepackage{graphicx}  
\usepackage{physics}
\usepackage[caption=false]{subfig}
\usepackage{amsmath, amssymb}
\usepackage{bm} 
\usepackage{tabularx}
\usepackage{orcidlink}
\usepackage[export]{adjustbox}

\usepackage[ruled, linesnumbered, titlenumbered, vlined]{algorithm2e}

\SetCommentSty{mycommfont}

\usepackage{comment}
\usepackage{bm}
\usepackage{soul}

\usepackage{tikz}
\usetikzlibrary{
	matrix, 
	quantikz2,
}

\usepackage{hyperref}
\usepackage[capitalise]{cleveref}
\crefname{section}{Sec.}{Secs.}
\Crefname{section}{Sec.}{Secs.}
\crefname{subsection}{Sec.}{Secs.}
\Crefname{subsection}{Sec.}{Secs.}
\crefname{appendix}{App.}{Apps.}
\Crefname{appendix}{App.}{Apps.}
\crefname{subappendix}{App.}{Apps.}
\Crefname{subappendix}{App.}{Apps.}
\hypersetup{
colorlinks=true,
urlcolor= magenta,
citecolor=blue,
linkcolor=blue,
}

\newcommand{\lind}{\mathcal{L}}
\newcommand{\dt}{\partial_t}
\newcommand{\da}{\nabla_{\bm{\alpha}}}

\newcommand{\dajk}{\partial_{\alpha_{jk}}}

\newcommand{\superI}{\mathcal{A}}

\newcommand{\heis}[1]{#1^{(\mathrm{H})}}

\newcommand{\partialder}[2]{\frac{\partial #1}{\partial #2}}
\newcommand{\alphak}{\alpha_k}

\newcommand{\bigO}[1]{\mathcal{O}\qty(#1)}

\newcommand{\maintextsection}[1]{\paragraph*{#1 ---}}

\begin{document}

\title{Analytical Series Expansion for Efficient Gradient Evaluation\texorpdfstring{\\}{ }in Multi-Qubit Optimal Control}

\author{Ashutosh Mishra\,\orcidlink{0009-0008-4432-0408}}
\email{a.mishra@fz-juelich.de}
\affiliation{Institute for Quantum Computing Analytics (PGI-12) \\ Forschungszentrum J\"ulich, 52425 J\"ulich, Germany}
\affiliation{Theoretical Physics, Universit\"at des Saarlandes, 66123 Saarbr\"ucken, Germany}

\author{Elena Lupo\,\orcidlink{0000-0002-7019-2963}}
\affiliation{Institute for Quantum Computing Analytics (PGI-12) \\ Forschungszentrum J\"ulich, 52425 J\"ulich, Germany}

\author{Frank K. Wilhelm\,\orcidlink{0000-0003-1034-8476}}
\affiliation{Institute for Quantum Computing Analytics (PGI-12) \\ Forschungszentrum J\"ulich, 52425 J\"ulich, Germany}
\affiliation{Theoretical Physics, Universit\"at des Saarlandes, 66123 Saarbr\"ucken, Germany}

\author{Alessandro Ciani\,\orcidlink{0000-0002-8707-0532}}
\affiliation{Institute for Quantum Computing Analytics (PGI-12) \\ Forschungszentrum J\"ulich, 52425 J\"ulich, Germany}


\begin{abstract}
Gradient-based quantum optimal control and the theory of operator evolution are rarely discussed together. In this Letter, we bridge this gap by showing that the gradient of the time-evolved propagator involves the Heisenberg evolution of local operators. We present a unifying framework by deriving from first principles the formal solution for the gradient under an arbitrary pulse parameterization, and show that the same construction extends to derivatives of any order. We obtain a series of nested, time-independent commutators weighted by time-dependent scalar coefficients. The commutators are thus computed once, and each optimization step only updates the scalars. The method is particularly suited for simulating optimal control tasks in quantum systems with local interactions, which is a common situation in large multi-qubit platforms. In this setting, light-cone arguments and sparse-Pauli truncation heuristics may be used to keep the number of relevant terms in the series small.
We compare the computational cost required for the series with the Gradient Optimization of Analytic conTrols (GOAT) method, and, focusing on the problem of preparation of a GHZ state, demonstrate more than an order of magnitude speedup for a qubit ladder and a chain geometry. 

\end{abstract}

\maketitle

With the increasing size of quantum processors, precise calibration and control of qubits is essential to achieve high fidelities, while taking into account the effects of spurious interactions with spectator qubits. 
One potential approach to tackle this involves creating an accurate classical model, including spectator qubits, to find pulses to mitigate such errors. 
Quantum optimal control (QOC) techniques \cite{koch2022, glaser2015, Brif_2010} have been promising for the design of pulses to perform single and two-qubit gates \cite{sporl_optimal_2007, egger_adaptive_2014, machnes2018, cykiert2024, george2025}, target state preparation \cite{arbitrary_rojan-quantum-state_2014, 
Heeres2017, reich_optimal_2013, malis_local_2019}, qubit readout with high fidelity \cite{mishra2025, jerger2024, gautier2025}, and address problems such as frequency crowding \cite{schutjens_single-qubit_2013}, leakage reduction \cite{motzoiSimplePulsesElimination2009, werninghaus_leakage_2021, hyyppaReducingLeakageSingleQubit2024}, and crosstalk \cite{sheldon_procedure_2016, wang_control_2022}. 
Among the various QOC techniques, gradient-based methods \cite{reich2012, khaneja2005, defouquieres2011, machnes2018} are useful in finding optimal parameters in a more efficient way than gradient-free methods, as the information about the variational change of fidelity landscape can significantly reduce the number of iterations required for the optimization \cite{machnes2018}. 

Although the current QOC methods work well for systems with a few qubits, extending these to large Hilbert spaces or open quantum systems introduces significant computational overhead. Recent works have addressed this problem, focusing on systems of superconducting qubits coupled to a cavity mode~\cite{Lu_2024, gautier2025, boutin2017resonatorreset}.
Classically simulating and optimizing these models involve calculating the time-ordered propagator --- unitary operator for closed systems and Lindblad superoperator for open systems --- as well as their derivatives. 
However, due to time-ordering, computing the derivatives is particularly challenging for any general pulse ansatz.
QOC methods, like GRAPE \cite{khaneja2005} and Krotov's method \cite{krotov2019Goerz}, assume a piecewise-constant pulse ansatz to bypass these limitations. On the other hand, methods like GOAT \cite{machnes2018} compute the gradient by co-propagating a set of derivatives along with the propagator. This either restricts the user to the choice of specific parametrization, which may not be ideal for the problem \cite{pagano_rembold2024}, or significantly increases the computational cost for performing the optimization. 

In this Letter, we present a unifying framework for QOC and derive, from first principles, the formal solution for the gradient of the propagator for arbitrary pulse parameterizations. 
The approach can also be readily extended to open systems, and further, to obtain formal solutions for any $n^\text{th}$-order derivative of the propagator. 
Moreover, we derive a series expansion from the formal solution, and, using locality constraints, show that the series can be efficiently computed for optimization of large Hilbert spaces.
We compare the computational cost required for computing the gradient with the  series and the GOAT method, and show an order of magnitude speedup. Finally, we apply the method to the preparation of GHZ states in multi-qubit systems with a chain and a ladder geometry. We show that the method can be used to obtain non-trivial pulses with hundreds of parameters that yield high-fidelity GHZ states. A similar problem for the generation of $\mathrm{W}$ states has been recently studied experimentally in a superconducting chip in Ref.~\cite{romeiro2026}, while Ref.~\cite{muratori2025} provided a numerical study of the preparation of $\mathrm{W}$ and Dicke states with $\mathrm{X}, \mathrm{Y},$ and $\mathrm{ZZ}$ control, which is again similar to our model of GHZ state preparation.  

Our method highlights that computing the gradients of typical loss functions in QOC involves evaluating the evolution of gradients of the Hamiltonian terms in the Heisenberg picture. Thus, the problem is mapped to the study of operator evolution which is of central interest in several fields, such as the study of the out-of-time-order correlators~\cite{abaninObservationConstructiveInterference2025}, the theory of operator spreading \cite{chen2023, parker_universal_2019}, scrambling dynamics and quantum chaos \cite{xu_locality_2019}. 
Methods like Pauli propagation \cite{rudolph2025paulipropagationcomputationalframework}, tensor networks \cite{orus_practical_2014, biamonte_tensor_2017}, sparse Pauli dynamics \cite{begusic2025realtimeoperatorevolution, begusic_fast_2024}, etc., have been popular for the study of operator evolution. By connecting QOC with operator evolution, we use the inherent light cone structure \cite{lieb_finite_1972, kliesch_lieb-robinson_2014} to efficiently approximate the gradient. Further, we borrow concepts from sparse Pauli dynamics and implement common heuristic truncation strategies~\cite{rudolph2025paulipropagationcomputationalframework} to reduce the number of terms needed to evaluate the gradients.

\maintextsection{QOC Problem}
A QOC problem can be described by a drift Hamiltonian $H_0$ and a set of control terms $H_m$, $m=1, \dots, M$. The full Hamiltonian takes the general form
\begin{align}
\label{eq:hqoc}
    H(\bm{\alpha}, t) = H_0 + \sum_{m = 1}^M u_m(\bm{\alpha}, t) H_m
\end{align}
with $\bm{\alpha}  =\begin{pmatrix}
    \alpha_{1}, & \alpha_{2}, & \dots, & \alpha_{N_{\alpha}}
\end{pmatrix}$ representing a list of $N_{\alpha}$ optimizable parameters associated with the control pulses $u_m$. Note that in typical situations, most of the pulses will depend only on a subset of parameters, but we leave a general $\bm{\alpha}$ dependence in \cref{eq:hqoc}.

In a QOC problem, the optimization aims at finding the optimal parameters $\bm{\alpha}$ that minimize a cost function $C$. The cost function typically takes the form $C= 1 - \mathcal{F}$, where $\mathcal{F}$ represents either the gate or state-transfer fidelity. Assuming a closed system, for the state-transfer fidelity $\mathcal{F} \equiv |\bra{\psi_f}\ket{\psi(t)}|^2$, the gradient of the cost function $\da{C}$ can be expressed as
\begin{align}
\label{eq: gradient of C}
\begin{split}
    \da{C}  &= - \da{\mathcal{F}} \\ &= -\bra{\psi_f}\da U(t)\ket{\psi(0)} \bra{\psi(t)}\ket{\psi_f} - \text{c.c}.,
\end{split}
\end{align}
where $\ket{\psi(t)} = U(t) \ket{\psi(0)}$ is the evolved state up to time $t$ and $\ket{\psi_f}$ the final, target state. Additionally, $U(t) \equiv U(t,0) $, i.e., the unitary propagator from the initial time $0$ to the final time $t$, with the notation that the initial time is assumed to be zero, unless explicitly stated otherwise.
A similar expression can be derived for the gate fidelity, where $\mathcal{F} = (1/\mathrm{dim}(U_f)^2)|\mathrm{Tr}\{U^\dagger_f U(t)\}|^2$, with $U_f$ being the final, target unitary. It is therefore important to be able to compute $\da U(t)$ efficiently, especially when the size of the Hilbert space increases.\\
Here we focus on the calculation of the gradient of $U(t)$ for applications to closed systems. An extension to the case of open systems can be found in \cref{Appendix Section: Formal solution for gradients open systems} of Supplemental material (SM). Also, we focus on the case where $u_m(\bm{\alpha}_m, t)$ can be expressed by analytical functions; however, the method can easily be extended for the case of piecewise-constant pulses.

\begin{figure}
    \centering
    \includegraphics{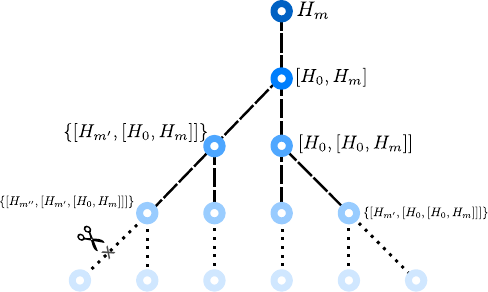}
    \caption{Schematics of the commutator-tree, when the drive operators $H_m$ are 1-local, for iteratively evaluating the nonzero terms of \cref{eq:seriesterms}. The terms in curly-brackets represent multiple terms with $m', m'' \neq 0$, and represent the commutation with drive operators. The terms at the same order, represented here by the same shade, can be computed in parallel from the previous order. Nodes at high order with small commutator norm or small coefficient values are pruned.}
    \label{Main text fig: commutator-tree}
\end{figure}

\maintextsection{Gradient computation via series expansion} Starting from the system of equations in the GOAT protocol \cite{machnes2018}, it can be shown that the components of the gradient of the propagator $\da U(t)$ for a closed system, represented by the Hamiltonian in \cref{eq:hqoc}, can be written as follows (see SM~\cref{Appendix Section: Formal solution for gradients closed system} for a detailed derivation):
\begin{multline}\label{eq:parderintegralform}
    \frac{\partial U(t)}{\partial \alphak} = -i U(t) \int_{0}^t d \tau \frac{\partial u_m (\tau)}{\partial \alpha_k} H_m^{(\mathrm{H})}(\tau) = \\
    \left [-i U(t) \int_{0}^t d \tau \frac{\partial u_m (\tau)}{\partial \alpha_k} H_m^{(\mathrm{H})}(\tau) U^{\dagger}(t) \right ] U(t),
\end{multline}
where the superscript $(\mathrm{H})$ denotes an operator in the Heisenberg picture, i.e., a general time-dependent operator $A(t)$ in the Schr\"{o}dinger picture becomes
\begin{equation}
A^{(\mathrm{H})}(t) = U^{\dagger}(t) A(t) U(t),
\end{equation}
in the Heisenberg picture.
The observation that the integrand in \cref{eq:parderintegralform} involves evolution of the derivative of the Hamiltonian in the Heisenberg picture is central to this work, and connects QOC to the theory of operator evolution. 
For simplicity, we have assumed that $\bm{\alpha}$ are not shared by multiple control Hamiltonians $H_m$. 
The results obtained here can easily be extended to the case with shared parameters. 
It should further be noted that the gradient of model parameters could also be computed using \cref{eq:parderintegralform} by substituting the integrand with $\heis{(\partial_{\alpha} H_0)}(\tau)$ for some model parameter $\alpha$. 
\cref{eq:parderintegralform} is exact and can be used, in principle, to compute the gradient for state-transfer problems, \cref{eq: gradient of C}, in terms of a forward propagation of the initial state: 
\begin{multline}
\label{eq:parderfidelity}
    \partialder{\mathcal{F}(t)}{\alphak} = \\ \bra{\psi_f} \left [-i U(t) \int_0^t d \tau \left( \frac{\partial u_m(\tau)} {\partial\alpha_k} \heis{H_m}(\tau)\right) U^{\dagger}(t) \right] \\ \times \rho(t) \ket{\psi_f}  +  \text{c.c.}
\end{multline}
where $\rho(t) = U(t)\ketbra{\psi(0)}{\psi(0)} U^{\dagger}(t)$ is the density matrix at time $t$. Although \cref{eq:parderintegralform} and \cref{eq:parderfidelity} provide a very simple way to compute the gradient, it is not efficient to compute the unitary propagator for every small time-step, especially for large Hilbert spaces where computing the propagator is expensive. 

To reduce the computational cost, we notice that the structure of \cref{eq:parderintegralform} involves evolving the drive operator in the Heisenberg picture, which evolves according to the Heisenberg equation $d_t\heis{H_m}(t) = i[\heis{H}(t), \heis{H_m}]$.
We integrate \cref{eq:parderintegralform} by parts and simplify it using the latter equation. By iterating this procedure $N$ times we obtain an approximated truncated series of the form (see \cref{Supp section: series exapnsion for unitary propagators} in SM for details):
\begin{equation}
\label{eq:gradientseriesexpansion}
    \partialder{U(t)}{\alphak} \simeq \qty(\partialder{U(t)}{\alphak})^{(N)}  = \left[\sum_{n = 0}^N \Theta_n(t) \right] U(t),
\end{equation}
with:
\begin{subequations}
\label{eq:seriesterms}
\begin{equation}
    \Theta_0(t) = -i \beta_{m}(t)H_m ,
\end{equation}
\begin{multline}
    \Theta_{n \ge 1}(t) = (-i)^{n+1} \sum_{\bm{m} \in [M]^n} \beta_{\bm{m}, m}^{(n)}(t) \\ \qquad\hspace{1cm} \times \left[H_{m^{(n)}}, \left[\dots, \left[H_{m^{(1)}}, H_m \right] \dots \right] \right],
\end{multline}
\end{subequations}
where $\bm{m} = \begin{pmatrix}
    m^{(n)}, & \dots, & m^{(1)}
\end{pmatrix} $ are $n$ integers from zero to $M$, and the coefficients $\beta_{\bm{m}, m}^{(n)}(t)$ are defined via the recursive formula
\begin{equation}\label{eq:betaintegralform}
    \beta_{\bm{m}, m}^{(n)}(t) = \int_0^t d\tau \, u_{m^{(n)}}(\tau) \, \beta_{\bm{m}^{n-1:1}, m}^{(n-1)}(\tau),
\end{equation}
where $\bm{m}^{n-1:1}$ denotes $\bm{m}$ without the first element $m^{(n)}$, and $u_0(t) = 1$. 
The time and parameter independence of the commutators in \cref{eq:seriesterms} suggests that the commutators need to be computed only once at the start of the simulation and stored, while, in each iteration of the optimization, only the coefficients $\beta_{\bm{m}, m}^{(n)}(t)$ need to be computed. \cref{Main text fig: commutator-tree} schematically shows the computation of the commutators involved in the series.
Since these integrals are easy to compute classically, this makes the calculation of the gradient much faster. We also emphasize here that no approximations have been used to derive \cref{eq:gradientseriesexpansion}. In fact, in the limit $N \rightarrow \infty$, and in its region of convergence, the series represents the exact gradient. 
Further, to show that \cref{eq:parderintegralform} is generally applicable to piecewise-constant parameterization, in \cref{Supp: connection with GRAPE} in the SM we reproduce the exact gradient of Ref.~\cite{defouquieres2011}. When comparing the structure of both series, it is clear that \cref{eq:gradientseriesexpansion} represents the generalization of the GRAPE gradient to continuous pulse parameterization.

For cases involving long times, we apply the composition rule of the propagator, and partition the time interval $[0, t]$ into a finer grid $[t_0=0, t_1) \cup [t_1, t_2) \cup \dots \cup [t_{n_I-1}, t_{n_I}=t]$, with $n_I$ the number of intervals. The gradient of the propagator can then be expressed as
\begin{align}\label{Main text eq: long time gradients}
    \da U(t) = U(t) \sum_j U^\dagger(t_j)  \left[ \sum_n \Theta_n(t_j, t_{j-1})\right]U(t_j),
\end{align}
where the series is computed only for the finer steps. We refer to \cref{Supp section: Gradients for long final times} in SM for a detailed derivation. Note that the series $\Theta_n(t_j, t_{j-1})$ can be computed in parallel for all the time steps.

Finally, we underscore the generality of our method for deriving the formal solution of the gradient, by deriving the form of the formal solution for any $n^\text{th}$ order derivative as (see SM~\cref{Supp section: formal solution for nth order derivative})
\begin{align}\label{eq:formalsolutionfornthderivative}
\begin{split}    
&\partial_\alpha^n U(t) = \sum_{(p, q, \dots) \in \mathcal{P}}  c_{(p, q, \dots)} (-i)^{\abs{(p, q, \dots)}}
\\ & \times \int_0^t d \tau^{(1)} \, U(t,\tau^{(1)}) \, \partial_\alpha^p H(\tau^{(1)})  
\\ & \dots \ \int_0^{\tau^{(j-1)}} \!\!\!\! d\tau^{(j)} \, U(\tau^{(j-1)}, \tau^{(j)}) \, \partial_\alpha^s H(\tau^{(j)}) \, U(\tau^{(j)}) \ ,
\end{split}
\end{align}
where $p, q, ...$ are positive integers that add up to $n$ $\big( p+q+r+s+... = n\big)$, i.e., the composition of the integer $n$; $\mathcal{P}$ is the set of all compositions of $n$, and $\abs{(p, q, \dots)}$ is the length of the ordered composition $(p, q, \dots)$.
The coefficients are given by $c_{(p, q, \dots)} = n!/(p! q! r! \dots 1)$, and the formal solution contains $2^{n-1}$ terms.

\begin{figure}
    \centering    \includegraphics{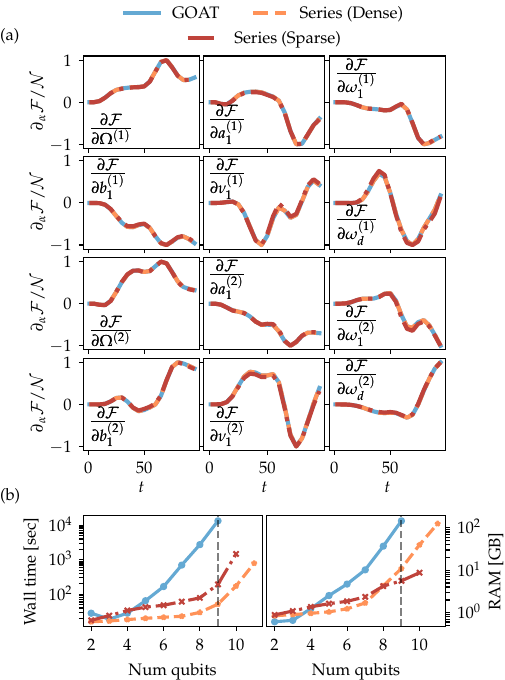}
    \caption{Normalized derivatives of the fidelity, given by the state-overlap $(\mathcal{F} = |\bra{\psi_f}\ket{\psi(t)}|^2)$, with respect to pulse parameters --- amplitude $(\Omega)$, Fourier coefficients $(a^{(i)}_j, b^{(i)}_j)$, Fourier frequency $(\omega^{(i)}_j, \nu^{(i)}_j)$, drive frequency $(\omega_d^{(i)})$ --- of a CRAB envelope along with sinusoidal oscillations. 
    (a) Comparison of normalized derivatives for two qubits computed using GOAT (blue), and derivatives by the series expansion using dense- (orange) and sparse-matrices (red).
    Here $\mathcal{N}$ represents the maximum absolute value of the gradient used for normalization.
    (b) Comparison of wall time (left) and peak memory usage (right) for the evaluation of the derivatives while varying the number of qubits in the chain with $12$ optimizable parameters. The plot is constructed for 9, 10, and 11 qubits for GOAT, sparse series and dense series respectively.
    The series expansion shows more than an order of magnitude improvement over GOAT in both time and memory requirements.}
    \label{Main text fig: comparision of gradients}
\end{figure}

\maintextsection{Locality}

In what follows, we consider the total Hamiltonian to have a graph structure, and composed of subsystems whose interactions can be defined on a graph $G = (V, E)$:
\begin{equation}
    \label{eq:graphHamiltonian}
    H(t) = \sum_{u \in V} H^{(u)}(t) + \sum_{(u,v) \in E} H^{(uv)}(t),
\end{equation}
where $H^{(u)}(t)$ denotes a Hamiltonian term acting non-trivially only on the subsystem vertex $u \in V$, while $H^{(uv)}(t)$  acts non-trivially only on the subsystems at edge $(u, v) \in E$. 
Thus, the total Hamiltonian in \cref{eq:graphHamiltonian} is a two-local Hamiltonian, i.e., with interactions acting on at most two subsystems. More generally, the Hamiltonian can be considered to be a $k$-local Hamiltonian -- a Hamiltonian that can be written as the sum of terms that act non-trivially on at most $k$ subsystems~\cite{kitaev_book}. We also consider the control Hamiltonians $H_m$ one- or two-local, which is a standard assumption in many quantum optimal control problems \cite{jandura2022time, Yan2018TunableCoupling, Perret2024Preparation}. 

By imposing the graph structure on the Hamiltonian, the number of terms in the summation in \cref{eq:seriesterms} can be reduced. For instance, only nearest-neighbor terms would contribute to the second-order term in \cref{eq:gradientseriesexpansion}, as the rest of the commutators are zero. 
For sparse connectivity, e.g., 1D chain, this drastically reduces the terms in the series.
Truncation error bound and convergence criteria for the series can be derived by bounding the number of nonzero nested commutator terms. The truncation number $N^*$ is found to scale logarithmically with the inverse of the target accuracy $\epsilon$ for short final time t,
\begin{equation}
\label{eq:series_truncation_main}
    N^* \sim \frac{\ln(1/\epsilon)}{\ln\big(1/(4n_dt)\big)},
\end{equation}
where $n_d$ is the maximum number of connections per vertex in \cref{eq:graphHamiltonian}. We refer to \cref{Supp section: Error bound on series truncation} for a detailed derivation. Further, the truncation error for long final time can be drastically reduced by using \cref{Main text eq: long time gradients}, and the series can be approximated using terms up to a few order for the shorter time intervals. 

\cref{eq:gradientseriesexpansion} involves computing one propagator $U(t)$ which, under a geometrically local Hamiltonian, can be computed using the Magnus expansion \cite{blanes2009, sharma2024hamiltoniansimulationinteractionpicture}. Locality also allows us to bound the support of the series in \cref{eq:gradientseriesexpansion} owing to Lieb-Robinson bounds, as we show in \cref{sec:localityandliebrobinson} in the SM. It should,  however, be noted that this analytical bound is not directly used in the numerical examples we show later, and only motivates the truncation of the series theoretically.

\maintextsection{Numerical computation of the series}
During an optimization process, the gradient with respect to the optimization parameters is computed using Algorithm~\ref{Algo1:Gateoptimisationusinggradientseries} (see SM~\cref{sec:numericalimplementation} for details of the implementation). At the start of the optimization, relevant commutators are iteratively computed and stored as a commutator-tree, as sketched in \cref{Main text fig: commutator-tree}. For the sparse implementation, the commutators are stored as Pauli strings. 
In each subsequent iteration only an analogous tree for the coefficients $(\beta_{\bm{m},m}^{(n)}(t))$ is computed, which are then multiplied with the commutators and summed over to obtain the value of the series. Furthermore, locality-based and heuristic metrics similar to sparse Pauli dynamics \cite{rudolph2025paulipropagationcomputationalframework, begusic2025realtimeoperatorevolution}, are used to efficiently choose terms of the series.

\begin{algorithm}[t]\label{Algo1:Gateoptimisationusinggradientseries}
    \caption{Gate optimization using \cref{eq:gradientseriesexpansion} and \cref{Main text eq: long time gradients}}
    \tcp{Compute indices for commutator-tree}
    $T_\text{index}^m \gets \{(m,), \, (l,m), \, (k,l,m), \, \dots \, \}$ \;
    
    \tcp{Precompute commutators from $T_\text{index}^m$}
    $T^m_\text{comm} \gets \{ H_m, \, [H_l, H_m], \, [H_k, [H_l, H_m]], \, \dots \, \}$ \;

    \tcp{Divide time grid into a few sections}
    $ t_{n} \gets 0 $\;
    \For{$t_{n+1} \in \{t_0, t_1, t_2, \dots t_{\mathrm{final}}\}$}{
        \tcp{Using Magnus expansion}
        $U(t_{n+1}) \gets e^{\Omega(t_{n+1}, t_{n})} U(t_{n})$;\\
        \tcp{Compute $\beta$s as tree from $T^m_\text{index}$}
        $T^m_\beta \gets \{ \beta_{m}(t_{n+1}, t_{n}), \, \beta_{l,m}(.), \, \beta_{(k, l), m}(.), \, \dots \, \}$\;
        $\da \gets \da + U^\dagger(t_{n+1}) \qty(\sum T^m_\text{comm} \, T^m_\beta) U(t_{n+1}) $\;    
        $t_{n} \gets t_{n+1}$\;
    }
    $\da \gets U(t_{\mathrm{final}}) \da$\;
    \Return $\da$
\end{algorithm}

\begin{figure}
    \centering
    \includegraphics{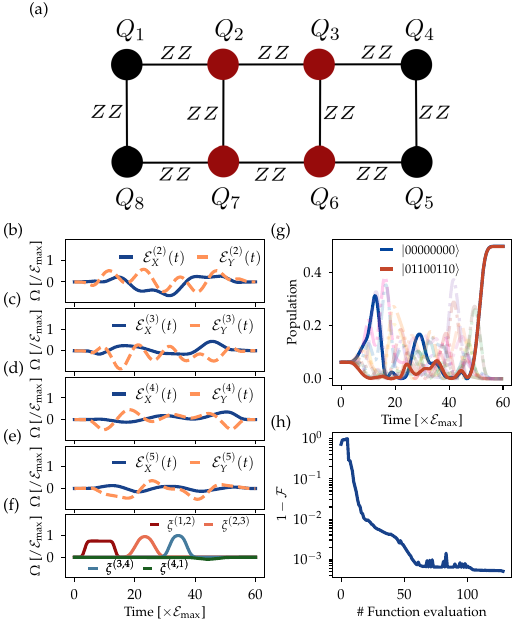}
      \caption[GHZ state-preparation on a 2D qubit ladder]{GHZ state-preparation on a 2D qubit ladder. (a) Schematics of a model of 8 qubits in a ladder geometry, with the aim to construct a 4-qubit GHZ state on center qubits $Q_2$, $Q_3$, $Q_6$, and $Q_7$ denoted in red by using the tunable $\mathrm{ZZ}$ interactions, in the presence of stray $\mathrm{ZZ}$ interactions on all edges. Here, qubits $Q_1$, $Q_4$, $Q_5$ and $Q_8$, shown in black, act as spectator qubits. (b-e) The optimized pulse shapes for the local $\mathrm{X}, \mathrm{Y}$ drives and (f) the optimized pulse shapes for the tunable $\mathrm{ZZ}$ interactions between the qubits, with a total of 264 optimizable parameters. (g) Corresponding dynamics, highlighting only the computational states relevant to the target GHZ state. (h) Infidelity for each function evaluation during the L-BFGS-B optimization, starting from a pulse optimized without stray $\mathrm{ZZ}$ interactions. }
    \label{fig:modelsystem2dladder}
\end{figure}

To validate our method, we consider a 1D chain of qubits with drive on the first and second qubits, with the aim to prepare a Bell state in the presence of spectator qubits. We compare the gradient of the fidelity computed using the series, \cref{eq:gradientseriesexpansion}, with the exact gradient computed using co-propagating the derivatives as in the GOAT method \cite{machnes2018}.  
\cref{Main text fig: comparision of gradients}(a) shows a comparison of gradient of a two-qubit system with a CRAB envelope \cite{muller2022one, doria2011optimalcontrol, caneva2011choppedrandom} as pulse ansatz, including sinusoidal oscillations, details of which can be found in \cref{sec:latticesimulation} of the SM. 
Moreover, \cref{Main text fig: comparision of gradients}(b) compares the wall time and the memory required for both methods as a function of the number of qubits, and demonstrates that the series expansion provides more than an order of magnitude speedup especially for large system sizes. 
Finally, to further verify the accuracy of our method and its applicability in optimal control tasks, we consider a model system of a 2D ladder geometry of qubits with stray $\mathrm{ZZ}$ interaction on all edges; with local $\mathrm{X}$, $\mathrm{Y}$ drives and tunable $\mathrm{ZZ}$ couplings as controls, as highlighted in \cref{fig:modelsystem2dladder}(a). The goal is to prepare a 4-qubit GHZ state on the four center qubits. \cref{fig:modelsystem2dladder}(g) demonstrates the dynamics of the system under the optimized pulses shown in \cref{fig:modelsystem2dladder}(b-f), with a fidelity of 99.95\%. \cref{fig:modelsystem2dladder}(h) shows the L-BFGS-B \cite{byrdLimitedMemoryAlgorithm1995} optimization process for each function evaluation, demonstrating a quick decrease in infidelity.

\maintextsection{Applications}
In addition to the standard QOC, such an analysis of derivative of the propagator has applications to other fields of interest such as computation of quantum Fisher information \cite{pangQuantumMetrologyGeneral2014a, pang_optimal_2017, huControlIncompatibilityMultiparameter2024}, estimating the quantum control landscape \cite{russell2017control, beato_towards_2024, chakrabarti_quantum_2007, de_fouquieres_closer_2013, wu_singularities_2012}, and robust QOC \cite{poggi_universally_2024, nelson_designing_2023}.
Furthermore, this work connects QOC with study of operator evolution, thus extending tools like tensor network and sparse Pauli propagation to the optimization of quantum systems.

\maintextsection{Conclusion}
We developed a unified framework for quantum optimal control to compute the gradient of time-ordered unitary propagator under arbitrary pulse ans\"{a}tze. We derived the formal solution for the gradient, and showed that this method can be readily extended to derive the formal solution for higher-order derivatives and open quantum systems. Moreover, we derived a series expansion that can be efficiently computed for locally-interacting multi-partite quantum systems. With a numerical implementation, we showed more than an order of magnitude speedup compared to the GOAT method.
Finally, with this work, we motivate the development of methods to perform operator and state evolution using sparse representation under a time-dependent Hamiltonian to scale QOC to even larger quantum systems. 

\maintextsection{Data availability} 
We provide a Python based repository for this work - \url{https://jugit.fz-juelich.de/pgi-12-external/qfc/applications/analyticGradient}. 
Our code is based on the open source Python package ParaQeet \cite{paraqeet}, which A. M. and A. C. co-developed.

\maintextsection{Acknowledgements}    
A. M. acknowledges funding from the German Federal Ministry of Education and Research (BMBF) under the program ``German Quantum Computer based on Superconducting Qubits'' (Project GeQCoS) under the contract number 13N15680. E. L. acknowledges funding from the the German Federal Ministry of Education and Research (BMBF) under the program ``Munich Quantum Valley Quantencomputer Demonstratoren – Supraleitende Qubits'' (Project MUNIQC-SC) under the contract number 13N16190.

\let\oldaddcontentsline\addcontentsline
\renewcommand{\addcontentsline}[3]{}
\bibliography{bibliography}
\let\addcontentsline\oldaddcontentsline

\clearpage
\onecolumngrid
{\center \bf \large Supplemental Material for ``Analytical Series Expansion for Efficient Gradient Evaluation in Multi-Qubit Optimal Control"\\}
{\center Ashutosh Mishra,\textsuperscript{1,\,2} Elena Lupo,\textsuperscript{1} Frank K. Wilhelm,\textsuperscript{1,\,2} and Alessandro Ciani\textsuperscript{1}\\\vspace*{-0.25cm}}
{\center \small \textsuperscript{1}\textit{Institute for Quantum Computing Analytics (PGI-12) \\ Forschungszentrum J\"ulich, 52425 J\"ulich, Germany}\\\vspace*{-0.3cm}}
{\center \small \textsuperscript{2}\textit{Theoretical Physics, Universit\"at des Saarlandes, 66123 Saarbr\"ucken, Germany}\\\vspace*{1cm}}

\setcounter{equation}{0}
\setcounter{figure}{0}
\setcounter{table}{0}
\setcounter{page}{1}
\setcounter{section}{0}

\makeatletter
\renewcommand{\theequation}{S\arabic{equation}}
\renewcommand{\theHequation}{S\arabic{equation}}
\renewcommand{\thefigure}{S\arabic{figure}}
\renewcommand{\theHfigure}{S\arabic{figure}} 
\renewcommand{\thesection}{S\arabic{section}}
\renewcommand{\theHsection}{S\arabic{section}}
\renewcommand{\thetable}{S\arabic{table}}
\renewcommand{\theHtable}{S\arabic{table}}

\tableofcontents

~\newline

The Supplemental Material contains detailed derivations and numerical implementations, and is organized as follows. 
\cref{Appendix Section: Formal solution for gradients} contains a first-principles derivation of the formal solution for the gradient of both unitary and, more generally, quantum channel propagators.
\cref{Appendix section: series expansion} extends the formal solution to compute a series expansion for the gradient of the unitary propagator, with a discussion on gradient computation for long final times. We derive convergence criteria and truncation error bound for the series for short and long final time.
\cref{sec:localityandliebrobinson} explains how the Lieb-Robinson bound can be used to motivate truncation of the series in locally-interacting multi-partite systems. Further, this section motivates the use of Magnus expansion for computing the propagator.
\cref{sec:connectionwithotherqoc} includes the derivation of exact gradient for piecewise-constant pulses, and the derivation of formal solution for higher-order derivatives of the propagator.
Finally, \cref{sec:numericalimplementation} discusses the current code implementation used for computing the gradient, including sparse implementation and model systems used in the numerical examples discussed in the main text.

Equations and figures that are not labeled with the symbol ``S" refer to those in the main text.

\section{Formal solution for gradient of propagator}\label{Appendix Section: Formal solution for gradients}
In this section, we derive from first principles the formal solution for the gradient of propagator generated by a Hamiltonian or a Lindbladian for closed and open systems, respectively.

\subsection{Closed system}
\label{Appendix Section: Formal solution for gradients closed system}

We start from the Schr\"odinger equation for the unitary propagator, generated by the Hamiltonian parameterized by some parameters $\bm{\alpha}$ ($\hbar =1$)
\begin{align}
\label{eq:schrodinger}
    \partialder{}{t} U(\bm{\alpha}; t) = -i H(\bm{\alpha}, t) U(\bm{\alpha}; t).
\end{align}
In this text, we denote the unitary propagator as $U(t) \equiv U(t, 0)$, i.e., with initial time as zero, unless explicitly stated otherwise.
We compute the gradient of the propagator with respect to the parameters $\bm{\alpha}$ by differentiating  both sides of the above equation, and switching the order of derivatives for the term on the left
\begin{align}\label{eq:goatgrad}
    \partialder{}{t} (\da U) = -i (\da H) U -i H (\da U).
\end{align}
For the sake of compactness, the explicit dependence on $t$ and $\bm{\alpha}$ have been suppressed in the equations below.
\cref{eq:schrodinger} and \cref{eq:goatgrad} form a set of coupled differential equation, expressed in the matrix form by
\begin{align}\label{eq:goatmatrix}
    \partialder{}{t} \begin{pmatrix} U \\ \da U \end{pmatrix} = -i
    \begin{pmatrix} H & 0 \\ \da H & H \end{pmatrix} \begin{pmatrix} U \\ \da U \end{pmatrix},
\end{align}
as expressed in the Gradient Optimization of Analytic conTrols (GOAT) method~\cite{machnes2018}. These equations can be solved numerically by co-propagating the gradient either by ODE solvers, or by exponentiating the auxiliary matrix.
Here we find the formal solution to the gradient of the propagator by solving \cref{eq:goatgrad} analytically.

By multiplying \cref{eq:goatgrad} from the left by $U^{\dagger}$ and using the Hermitian conjugate of the Schr\"{o}dinger equation \cref{eq:schrodinger}, we obtain
\begin{align}\label{eq:gradintermediate}
    \partialder{}{t} (U^\dagger \da U) = -i U^\dagger (\da H) U.
\end{align}
Integrating both sides of \cref{eq:gradintermediate}, we obtain that the gradient can be expressed as
\begin{align}
\label{eq:gradintegralform}
    (\da U) (t) = -i U(t) \int_0^t d \tau \, U^\dagger(\tau) \, (\da H)(\tau) \, U(\tau),
\end{align}
where we have assumed that the gradient is zero at time $t=0$.
The form of this equation resembles Duhamel's formula \cite{wilcox1967exponential, nachtergaele2019quasilocality}, and the derivative of an exponential map \cite{hall_book}. 
This form of the derivative of the propagator exists in literature \cite{HO2006226, pang_optimal_2017}; here we derive it from first principles and demonstrate that this same form works for any time-dependent Hamiltonian. Further, in \cref{Supp section: formal solution for nth order derivative},  we show that from the first principles any $n^\text{th}$-order derivative of the propagator can be written in a similar integral form.

The integrand in \cref{eq:gradintegralform} represents the operator $\da H(t)$ in the Heisenberg picture. This observation is central to our work, and is used in \cref{Appendix section: series expansion} to derive a series expansion.
We assume that a certain parameter $\alpha_k$ modifies only one pulse $u_m$, which mathematically means $\partial u_{m'}/\partial \alpha_k =0$, $\forall m' \neq m$. Note that considering a Hamiltonian as in \cref{eq:hqoc}, this is without loss of generality, since we are leaving the definition of the $H_m$ arbitrary. With this assumption, using \cref{eq:gradintegralform} the partial derivative of the unitary with respect to $\alpha_k$ can be written as in \cref{eq:parderintegralform}.

\subsection{Open systems} 
\label{Appendix Section: Formal solution for gradients open systems}

Generalizing the above, we show that the gradient of a quantum channel with respect to parameters of the Lindblad superoperator can similarly be derived by starting from the master equation. 
For some quantum channel $\Lambda(t)$, generated by the Lindbladian $\lind(\bm{\alpha}, t)$, where $\bm{\alpha}$ are the parameters of the Hamiltonian or the Lindbladian, the evolution of the quantum channel is given by the master equation
\begin{align}\label{eq: EOM quantum channel}
    \partialder{}{t} \Lambda(\bm{\alpha}; t) = \lind(\bm{\alpha}, t) \Lambda(\bm{\alpha}; t).
\end{align} 
Similar to the closed system, differentiating with respect to $\bm{\alpha}$ and switching the order of derivatives for the term on the left produces a coupled differential equation for the gradient of the quantum channel
\begin{align}\label{Supp Eq: gradient of quantum channel differential form}
    \partialder{}{t}(\da \Lambda) = (\da \lind) \Lambda + \lind (\da \Lambda).
\end{align}
This can be converted to a homogeneous differential equation by defining an \emph{auxiliary map} $(\superI)$ as
\begin{align}
    \superI(\bm{\alpha}; t) = \mathcal{T}_{\rightarrow} \exp{-\int_0^t d\tau \, \lind(\bm{\alpha},\tau)}
\end{align}
with $\mathcal{T}_{\rightarrow}$ representing the inverse time-ordering. It satisfies the differential equation
\begin{align}
    \frac{\partial \superI }{\partial t} = - \superI \ \lind.
\end{align}
We call this an auxiliary map as it is analogous to $U^\dagger$ for the closed case, and a true inverse of a quantum channel does not always exist, so the auxiliary map may not be a physical quantum channel.

Left multiplying both sides of the \cref{Supp Eq: gradient of quantum channel differential form} with the auxiliary map, produces a homogeneous differential equation
\begin{align}
    \partialder{}{t} ( \superI \ \da \Lambda) = \superI (\da \lind) \Lambda.
\end{align}
Integrating both sides and left-multiplying with the inverse of the auxiliary map, we obtain the formal solution for the gradient of a quantum channel
\begin{align}\label{Supp Eq: Gradient of channels integral form}
    \da \Lambda(t) = \superI^{-1}(t) \int_0^t d\tau \, \superI(\tau) (\da \lind)(\tau) \Lambda(\tau),
\end{align}
which is similar to \cref{eq:gradintegralform}. It should be noted that this equation is only valid if the auxiliary map is invertible.
In case the map is non-invertible, the inverse of the auxiliary map can be replaced by the Petz recovery map \cite{junge_universal_2016, petz_sufficient_1986, petz_sufficiency_1988}.

\subsection{Gradient for state-transfer problems}\label{Supp section: Gradients for state-transfer problems}
For state-preparation tasks, the cost function $(C)$ can be expressed as a function of state-overlaps $(\mathcal{F})$. Accordingly, the gradient of the cost function can be expressed as
\begin{align}
    \da C(\mathcal{F}) = \frac{\partial C(\mathcal{F})}{\partial \mathcal{F}} \da \mathcal{F}.
\end{align}
For a closed system, from \cref{eq:gradintegralform}, the gradient of the state-overlap can then be expressed in terms of two state-propagations -- forward propagation of the initial state $(\ket{\psi(t)} = U(t) \ket{\psi_0})$, and backward propagation of the target state  --
\begin{align}\label{Supp Eq: Gradient state-overlap closed system}
\begin{split}
    (\da \mathcal{F})(t) = -i \int_0^t d\tau \mel{\psi_f(\tau)}{\da H(\tau)}{\psi(\tau)}\, & \times \bra{\psi(t)}\ket{\psi_f}
    + \text{c.c.}
\end{split}
\end{align}
where $\mathcal{F}(t) = |\bra{\psi_f}\ket{\psi(t)}|^2$ and $\bra{\psi_f (\tau)} = \bra{\psi_f} U(t, \tau)$.

\cref{Supp Eq: Gradient state-overlap closed system} has a similar form as the gradient obtained from GRAPE \cite{khaneja2005}, but we emphasize that these equations are not tied to any specific pulse parameterization and are applicable equally to piecewise-constant or analytic pulse shapes. 
In \cref{Supp: connection with GRAPE} we demonstrate that the exact gradient from GRAPE \cite{defouquieres2011} can be reproduced by starting from \cref{eq:gradintegralform}.

\section{Series expansion}\label{Appendix section: series expansion}
In this section, we derive the series expansion of \cref{eq:gradintegralform} presented in the main text. This is helpful for problems involving gate optimization, or problems in which computing the unitary propagator is expensive. Using this we show that the number of matrix exponentiations can be drastically reduced; and later in \cref{suppsection:MagnusExpansion} we discuss how Magnus expansion can be used to efficiently compute the gradient. Later in this section, we also demonstrate how the series can be computed for long final time.

\subsection{Series expansion for gradient of unitary propagator}\label{Supp section: series exapnsion for unitary propagators}
The structure of \cref{eq:parderintegralform}  helps us to expand $\partial U(t) / \partial \alpha_k$ into a series by performing repeated integration by parts. The first integration by parts results in~\cite{Note1}
\begin{align}
\begin{split}
    \partialder{U(t)}{\alphak} =  & U(t) \Bigg\{-i\beta_{m}(t) \heis{H_m}(t) 
     +i \int_0^t d \tau \beta_{m}(\tau) \, \partialder{\heis{H_m}}{\tau}\, \Bigg\} \ ,
\end{split}
\end{align}
where we defined 
\begin{equation}
\label{eq:betam}
\beta_{m}(t) = \int_0^t d\tau \partialder{u_m(\tau)}{\alphak}.
\end{equation}
From now onwards we use the notation that $m = 0$ represents the contribution from the drift and $m \geq 1$ represents that of the drive Hamiltonian, with $u_{0}(t) = 1$. 
We also highlight that
\begin{equation}
    \heis{[A, B]}(t) = U^{\dagger}(t) [A, B] U(t) = [\heis{A}(t), \heis{B}(t)].
\end{equation}

The evolution of the $H_m$ in the Heisenberg picture, is governed by the Heisenberg equation of motion, which can be written as
\begin{align}
\label{eq:hkse}
\begin{split}
    \partialder{\heis{H_m}}{t} &= i \left[\heis{H}(t), \heis{H_m}(t) \right] =
    i \sum_{m'=0}^M u_{m'}(t)  \left[\heis{H_{m'}}(t), \heis{H_m}(t)\right].
    \end{split}
\end{align}
Thus the components of the gradient simplify to:
\begin{align}
\label{eq:gradseries1}
\begin{split}
    &\partialder{U(t)}{\alphak}  = U(t) \Bigg\{-i\beta_{m}(t) \heis{H_m}(t) - \sum_{m'=0}^M \int_0^t d\tau \beta_{m}(\tau) u_{m'}(\tau)  \left[\heis{H_{m'}}(\tau), \heis{H_m}(\tau) \right] \ 
    \Bigg\} \ .
\end{split}
\end{align}
Proceeding with an additional integration by parts, we obtain 
\begin{align}
\label{eq:du_int_by_part_second_order}
\begin{split}
   \partialder{U(t)}{\alphak} &=  U(t) \Bigg\{-i \beta_{m}(t) \heis{H_m}(t) - \sum_{m' = 0}^M \beta_{m',m}^{(1)}(t)  \left[\heis{H_{m'}} (t), \heis{H_m}(t) \right]  + \sum_{m' = 0}^M \int_0^t d\tau \beta_{m',m}^{(1)}(\tau) \, \partialder{([\heis{H_{m'}}, \heis{H_m} ])}{\tau} \Bigg\},
\end{split}
\end{align}
where we have introduced the notation
\begin{align}\label{eq:beta_second_order}
\begin{split}
   \beta_{m',m}^{(1)}(t) &= \int_0^t dt' \, u_{m'}(t') \beta_{m}(t') = \int_0^t dt' \, u_{m'}(t') \int_0^{t'} dt'' \, \partialder{u_m(t'')}{\alphak}
\end{split}
\end{align}
to represent the nested integral. We now keep on iterating this procedure. The derivative in the integrand of the final term in \cref{eq:du_int_by_part_second_order} can be obtained from the Heisenberg equation associated with the commutator $[\heis{H_{m'}}(t), \heis{H_m}(t)]$ and reads
\begin{align}\label{eq:heiscommhmmprime}
     \partialder{([\heis{H_{m'}}, \heis{H_m}])}{t} = i \left[\heis{H}(t), \left[\heis{H_{m'}}(t), \heis{H_m}(t)\right] \right] = i \sum_{m''=0}^M u_{m''}(t) \left[\heis{H_{m''}}(t), \left[\heis{H_{m'}}(t), \heis{H_m}(t)\right] \right].
\end{align}
Using this, \cref{eq:du_int_by_part_second_order} takes the form:
\begin{align}
\label{eq:du_int_by_part_third_order}
\begin{split}
     \partialder{U(t)}{\alphak} =  U(t) \Bigg\{ &-i \beta_{m}(t)\heis{H_m}(t) - \sum_{m'=0}^M \beta_{m',m}^{(1)}(t) \left[\heis{H_{m'}}(t), \heis{H_m}(t) \right] \\
    &+ i\sum_{m', m''=0}^M \int_0^t d \tau  u_{m''}(\tau) \beta_{m',m}^{(1)}(\tau) \left[\heis{H_{m''}}(\tau), \left[\heis{H_{m'}}(\tau), \heis{H_m}(\tau)\right]  \right]\Bigg\}.
\end{split}
\end{align}
Proceeding in this way up to order $N$, we can approximate the derivative as in \cref{eq:gradientseriesexpansion}, 
with the $\Theta_n (t)$ given in \cref{eq:seriesterms}.
Note that we used $U(t) \heis{A}(t) = A(t) U(t)$, for any time-dependent operator $A(t)$, to express the commutators as stated in \cref{eq:seriesterms}. We further note that comparing with \cref{eq:parderintegralform}, we obtain
\begin{equation}
\label{eq:thetaseriestoint}
    \lim_{N \rightarrow + \infty} \left[\sum_{n = 0}^N \Theta_n(t) \right] = 
    -i U(t)  \int_{0}^t d \tau \frac{\partial u_m (\tau)}{\partial \alpha_k} H_m^{(\mathrm{H})}(\tau) U^{\dagger}(t).
\end{equation}

\subsection{Gradient for long final times}\label{Supp section: Gradients for long final times}
For long final time, the number of terms required in \cref{eq:gradientseriesexpansion}  to maintain a certain accuracy needs to be increased. Instead, one can use a trick to divide the computation into shorter intervals, and hence, keep fewer terms in the series. 
Specifically, using the composition rule of the propagator, and partitioning the time interval $[0, T]$ into a finer grid $[t_0=0, t_1) \cup [t_1, t_2) \cup \dots \cup [t_{n_I-1}, t_{n_I}=t]$, with $n_I$ the number of intervals, the propagator can then be expressed as
\begin{align}
    U(t) = U(t, t_{n_I-1}) U(t_{n_I-1}, t_{n_I-2})\cdots U(t_2,t_1) U(t_1,0) = \prod_{j=1}^{n_I} U(t_{j},t_{j-1}).
\end{align}

The gradient of the unitary propagator can then be computed by using the product formula
\begin{align}\label{eq:unitarycompositederivative}
\begin{split}
    \da U(t) &= \sum_j U(t, t_{j}) \, \big(\da U(t_{j}, t_{j-1})\big) \, U(t_{j-1}) \\
    &= U(t) \sum_j U^\dagger(t_{j}) \, \big(\da U(t_{j}, t_{j-1})\big) \, U(t_{j-1}).
\end{split}
\end{align}
The derivative of the unitary propagator can be replaced by \cref{eq:gradientseriesexpansion} 
\begin{align}
    \da U(t_{j}, t_{j-1}) = \left[\sum_n \Theta_n(t_j, t_{j-1})\right] \, U(t_{j}, t_{j-1}),
\end{align}
simplifying \cref{eq:unitarycompositederivative} to \cref{Main text eq: long time gradients}.

Although the gradient at the initial time of each slice is no longer zero, the series remains valid because the resulting constants of integration accumulate into a single term, which can be set to zero.
Note that the series $\Theta_n(t_j, t_{j-1})$ can be computed in parallel for all the time steps.
Finally, \cref{Main text eq: long time gradients} implies that for state transfer problems one can store the forward and backward propagation of the initial and target states in a sparse grid and compute the gradient similar to GRAPE.

\subsection{Error bound on series truncation}\label{Supp section: Error bound on series truncation}
In this section we estimate a convergence criterion for the series \cref{eq:gradientseriesexpansion}, together with its truncation bound. The latter is evaluated without assuming any graph geometry, and thus can be used for studying the generic behaviour of the series.

\subsubsection{Bound on the integral coefficient}
Assuming that $|u_m|\leq 1$, $\forall\ m\geq 1$, we can bound the coefficients $\beta_{\bm{m}, m}^{(n)}$ in \cref{eq:betaintegralform}  with
\begin{align}
    \begin{split}
        \big|\beta_{\bm{m}, m}^{(n)}(t) \big| \leq \int_0^t dt^{(n)} \int_0^{t^{(n)}} dt^{(n-1)} \dots \int_0^{t^{(1)}} dt' \left| \frac{\partial u_m(t')}{\partial\alpha_k} \right| =\frac{1}{n!} \int_0^t dt' \ (t-t')^{n} \left| \frac{\partial u_m(t')}{\partial\alpha_k} \right|,    
    \end{split}
\end{align}
where we have used the Cauchy formula for repeated integration. By applying a change of variable in the integral, $t' \to t-\tau$, we can argue that 
\begin{align}
\label{beta_bound_SM}
\big| \beta_{\bm{m}, m}^{(n)}(t) \big| &{}\leq \frac{1}{n!} \int_0^t d\tau \ \tau^{n} \left| \frac{\partial u_m(t-\tau)}{\partial\alpha_k} \right| \leq \max_{t'\in[0,t]}\left|\frac{\partial u_m(t')}{\partial\alpha_k}\right|\frac{1}{n!} \int_0^t \tau^{n}d\tau = c\,\frac{t^{n+1}}{(n+1)!} \equiv |\beta^{(n)}|,
\end{align}
with $c = \max_{t'\in[0,t]}|\partial u_m(t')/\partial\alpha_k|$. The full series $\sum_n\beta^{(n)}\|\mathrm{ad}_H^n(H_k)\|$ then converges provided the commutator norms grow no faster than $\|\mathrm{ad}_H^n(H_k)\|\lesssim (n+1)!\,r^{n+1}$, in which case the series is bounded by a geometric one, $\sum_n\beta^{(n)}\|\mathrm{ad}_H^n(H_k)\| \lesssim c\sum_n (rt)^{n+1}$, converging within the finite radius $t<t^* = 1/r$.

\subsubsection{Bound on the commutators for graph Hamiltonian with finite interaction range}
Let's consider a graph Hamiltonian,
$$ H_0 = \sum_z h_z H_z,$$
with $H_z$ a $k$-local operator acting on the subset $S_z = \mathrm{supp}\{H_z\}$, $\| H_z \| = 1$, and $|h_z| \leq 1$. The total Hamiltonian including the control term is
\begin{equation}
\label{hamiltonian_graph_total_sum_SM}
    H = \sum_z h_z H_z + \sum_{m} u_m(t) H_m = \sum_{X} h_X(t) H_X, \quad \text{with} \quad h_X(t) H_X = \begin{cases}
        h_z H_z,& \text{ for } X=S_z;\\
        u_m(t) H_m,& \text{ for } X=S_m.
    \end{cases}
\end{equation}
where $S_m = \mathrm{supp}\{H_m\}$ is the support of each control term $H_m$.\\
Using this notation, we can express the terms of the gradient's series (\ref{eq:gradientseriesexpansion}) as:
\begin{align}
    \Theta_{n \ge 1}(t) = (-i)^{n+1} \sum_{\bar{X}_n} \beta_{\bar{X}_n, m}^{(n)}(t)  \left[H_{X^{(n)}}, \left[\dots, \left[H_{X^{(1)}}, H_m \right] \dots \right] \right] = (-i)^{n+1} \sum_{\bar{X}_n} \beta_{\bar{X}_n, m}^{(n)}(t) C(\bar{X}_n),
\end{align}
with $C(\bar{X}_n) = [H_{X^{(n)}}, [H_{X^{(n-1)}},\dots [H_{X^{(1)}},H_m]]]$, the sum performed over chains of terms $\bar{X}_n = [X^{(n)}, X^{(n-1)},$ $ \dots, X^{(1)}]$ whose supports are ``connected" (see also, for instance, Ref.~\cite{wildalhambra2023}), i.e. 
$$ X^{(1)} \cap S_m \neq \emptyset \quad \wedge \quad  X^{(l)}\cap \left( \bigcup_{k=1}^{l-1} X^{(k)} \cup S_m \right) \neq \emptyset, \ \ l=2,\dots,n,$$
and:
\begin{equation}
\beta_{X^{(1)},m}^{(1)}(t) = \int_0^t dt_1\, h_{X^{(1)}}(t_1) \int_0^{t_1} d\tau \partialder{u_m(\tau)}{\alphak}, \qquad \beta_{\bar{X}_n,m}^{(n)}(t) = \int_0^t dt_n\, h_{X^{(n)}}(t_{n}) \beta_{\bar{X}_{n-1},m}^{(n-1)}(t_{n});
\end{equation}
terms with disconnected support do not contribute to the gradient.
We notice that in both cases where $X^{(k)} = S_{z^{(k)}}$ or $X^{(k)} = S_{m^{(k)}}(t)$, the coefficient $\beta_{\bar{X}_n,m}^{(n)}(t)$ is bounded by the same quantity $|\beta^{(n)}|$ above, Eq.~(\ref{beta_bound_SM}). We can thus write:
\begin{align}
\label{bound_Theta_n_eq0_SM}
    \left\|\Theta_{n}(t)\right\| =&{}\ \bigg\| \sum_{\bar{X}_{n}} \beta_{\bar{X}_n,m}^{(n)}(t)  C(\bar{X}_{n}) \bigg\| \leq \big|\beta^{(n)} \big| \sum_{\bar{X}_{n}} \Big\|  C(\bar{X}_{n}) \Big\| \nonumber\\
    =&{}\ \big|\beta^{(n)} \big|\sum_{z^{(n)}} \sum_{\bar{X}_{n-1}} \left\| \left[H_{z^{(n)}}, C(\bar{X}_{n-1})\right] \right\| + \big|\beta^{(n)} \big| \sum_{m^{(n)}} \sum_{\bar{X}_{n-1}} \left\|  \left[H_{m^{(n)}}, C(\bar{X}_{n-1})\right] \right\| \nonumber\\
    \leq&{}\ 2\big|\beta^{(n)} \big|\sum_{z^{(n)}} \sum_{\bar{X}_{n-1}} \left\| C(\bar{X}_{n-1}) \right\| + 2\big|\beta^{(n)} \big| \sum_{m^{(n)}} \sum_{\bar{X}_{n-1}} \left\|  C(\bar{X}_{n-1}) \right\|,
\end{align}
with $S_{z^{(n)}}\cap \left[\bigcup_{k=1}^{n-1} X^{(k)} \cup S_m  \right] \neq \emptyset$ and $S_{m^{(n)}}\cap \left[\bigcup_{k=1}^{n-1} X^{(k)} \cup S_m  \right] \neq \emptyset$. 
The bound \cref{bound_Theta_n_eq0_SM} is determined only by $| \beta^{(n)}|$, and by counting the number of nonzero terms $C(\bar{X}_n)$. Since whenever $ S_{z^{(n)}}\supseteq S_{m^{(n)}}$, the support of the commutator $\left[H_{z^{(n)}}, C(\bar{X}_{n-1}) \right]$ includes the support of $\left[H_{m^{(n)}}, C(\bar{X}_{n-1}) \right]$, we can consider the bound over the summation of these terms contributing equally,
\begin{align}
\label{bound_Theta_n_eq1_SM}
    \left\|\Theta_{n}(t)\right\| 
    \leq&{}\ 4\big|\beta^{(n)} \big|\sum_{\tilde{X}^{(n)}} \sum_{\bar{X}_{n-1}} \left\| C(\bar{X}_{n-1}) \right\|,
\end{align}
where $\tilde{X}^{(n)}$ now indicates terms connected to $\bigcup_{k=1}^{n-1} X^{(k)} $, whose supports are not subset of each other. Essentially, this correspond to re-defining the terms $h_XH_X$ in \cref{hamiltonian_graph_total_sum_SM}. Re-iterating over $n$, the bound \cref{bound_Theta_n_eq1_SM} takes the form
\begin{align}
\label{bound_Theta_n_eq2_SM}
    \left\|\Theta_{n}(t)\right\| 
    \leq&{}\ 4^n \big|\beta^{(n)}\big| \|H_m\| \prod_{j=1}^n w_j,
\end{align}
with $w_j$ the number of non-zero commutators at iteration $j$, i.e. when $[H_{\tilde{X}^{(j)}}, C(\bar{X}_{j-1})] \neq 0$. Given that $H_{\tilde{X}}$ can be either an on-site term or a two-local connecting term (see \cref{eq:graphHamiltonian}), this quantity corresponds to the number $s_{j-1}$ of sites contained in the support of $C(\bar{X}_{j-1})$ plus the number of new connections, yielding
$$ w_j \leq s_{j-1} + s_{j-1}\,n_d - (s_{j-1}-1) = s_{j-1}\,n_d + 1,$$ 
with $n_d$ the maximum number of connection per site (coordination number), and $(s_{j-1}-1)$ subtracted because it counts the minimal number of shared edges in a connected chain of length $s_{j-1}$. To bound the number of sites $s_j$ at each iteration, we notice that if the Hamiltonian is $k$-local, a new commutator with $h_XH_X$ can at most increase the support by $k-1$ terms, until the total number of sites $N_s$ is reached. Thus, in our $2$-local case, we have that $s_j \leq \min(s_0 + j, N_s)$, with $s_0$ the number of initial sites in the support of $H_m$. Therefore we have
\begin{align}
\label{eq:Prod_tj_bound_SM}
\prod_{j=1}^n w_j \le&{}\ \prod_{j=0}^{n-1} (n_d\,s_j + 1) \leq
\begin{cases}
n_d^n\prod_{j=0}^{n-1} (s_0 + j + 1/n_d), & n \le n^*\\
n_d^n\prod_{j=0}^{n^*-1} (s_0 + j + 1/n_d) (N_s+1/n_d)^{n-n^*}, & n > n^*,
\end{cases} \nonumber\\
&{} =
\begin{cases}
n_d^n\dfrac{\Gamma(s_0+n+1/n_d)}{\Gamma(s_0+1/n_d)}, & n \le n^*\\
n_d^n\dfrac{\Gamma(N_s+1+1/n_d)}{\Gamma(s_0+1/n_d)} (N_s+1/n_d)^{n-n^*}, & n > n^*,
\end{cases}  \sim
\begin{cases}
n_d^n\,C^{n+s_0-1}_{s_0-1}\, n!, & n \le n^*\\
n_d^n\,[N_s!/(s_0-1)!] N_s^{n-n^*}, & n > n^*,
\end{cases}
\end{align}
with $n^* = N_s-s_0+1$ and the last approximation valid for $n\gg 1$.\\
For $n\leq n^*$,\,\,$\prod_{j=1}^n w_j \lesssim  C^{n+s_0-1}_{s_0-1}n_d^n\, n!$ and the bound in \cref{bound_Theta_n_eq2_SM} takes the form,
\begin{align}
\label{bound_Theta_n_final_SM}
    \left\|\Theta_{n}(t)\right\| 
    \leq&{}\  c\,\|H_m\| C^{n+s_0-1}_{s_0-1}\frac{[4n_dt]^{n + 1}}{(4n_d) (n+1)} \sim A\,n^{s_0-2}(4n_dt)^{n+1},
\end{align}
with $A \sim O(1)$. The contributing terms $C^{n+s_0-1}_{s_0-1}[4n_dt]^{n + 1}/(n+1)$ decrease for $4n_dt<1$ and $n > [s_0(4n_dt)-2]/(1-4n_dt)$, approximating the terms of a geometric series converging for $t < t^* = 1/[4n_d]$. For generic $|h_z|$ a factor $|h| = \max_z\{ |h_z|\}$ can be added to the denominator. The truncation error at order $N > [s_0(4n_dt)-2]/(1-4n_dt)$ is bounded by the geometric tail
\begin{align}
  \label{eq:truncationtail}
    \varepsilon_N(t) &{}= \Big\|\sum_{n=N+1}^\infty \Theta_n(t)\Big\| \lesssim c\,\|H_m\| \sum_{n=N+1}^\infty C^{n+s_0-1}_{s_0-1}[4n_dt]^{n+1}/(n+1) \nonumber\\
    &{}= c\,\|H_m\| C^{s_0+N}_{s_0-1}\frac{[4n_dt]^{N+2}}{N+2} \sum_{m=0}^\infty \frac{(s_0+m+N)!(N+2)!}{(s_0+N)!(m+N+2)!}[4n_dt]^{m} \nonumber\\
    &=c\,\|H_m\| C^{s_0+N}_{s_0-1}\frac{[4n_dt]^{N+2}}{N+2}  \sum_{m=0}^\infty \prod_{j=1}^{m}\Bigl(1+\frac{s_0-2}{N+j+2}\Bigr) [4n_dt]^{m} \nonumber\\
    &\leq
    \begin{cases}
    c\,\|H_m\| C^{s_0+N}_{s_0-1}(N+2)^{-1}[4n_dt]^{N+2} \sum_{m=0}^\infty [4n_dt]^{m}, & s_0 =1;\\
    c\,\|H_m\| C^{s_0+N}_{s_0-1}(N+2)^{-1}[4n_dt]^{N+2} \sum_{m=0}^\infty \Bigl(1+\frac{s_0-2}{N+3}\Bigr)^m [4n_dt]^{m}, & s_0 \geq 2;
    \end{cases}\nonumber\\
    & < A C^{s_0+N}_{s_0-1}\frac{[4n_dt]^{N+2}}{1 - q_N4n_dt},
\end{align}
with $q_N = 1$ for $s_0 = 1$ and $q_N = (N+s_0+1)/(N+3)$ otherwise. For $s_0 = 1$, the requirement of $\varepsilon_N(t)\le \epsilon$ for a target tolerance $\epsilon$ gives
\begin{equation}
\label{eq:Ntruncation}
     N \ge N^* = \frac{\ln(1/\epsilon) + \ln\!\big(A/(1-4n_dt)\big)}{\ln\big(1/(4n_dt)\big)} - 2 \;\sim\; \frac{\ln(1/\epsilon)}{\ln\big(1/(4n_dt)\big)},
\end{equation}
which is the bound presented in the main text, \cref{eq:series_truncation_main}.
Thus the number of terms grows only logarithmically in the inverse accuracy and is independent of the system size. For $4n_dt\to 0$, $N^* \to 0$, thus for $4n_dt\ll1$ only a limited number of terms $\Theta_n$ terms are needed, whereas as $4n_dt\to1$, $N^*$ diverges. It is safe to split the total evolution into steps $\Delta t < t^*$ chosen so that $4n_d\,\Delta t$ is comfortably below one.\\
For $n>n^*$,\,\,$\prod_{j=1}^n w_j \lesssim  \frac{N_s!}{(s_0-1)!N_s^{n^*}} (N_sn_d)^n$ and the bound in \cref{bound_Theta_n_eq2_SM} takes the form,
\begin{align}
\label{bound_Theta_n_final_Ns_SM}
    \left\|\Theta_{n}(t)\right\| 
    &{}\leq c\,\|H_m\| \frac{N_s^{s_0-1}}{(s_0-1)!}\frac{N_s!}{N_s^{N_s}} \frac{[4N_sn_dt]^{n+1}}{(4N_s n_d) (n+1)!} \lesssim A\,\frac{N_s^{s_0-3/2}}{e^{N_s}}  \frac{[4N_sn_dt]^{(n+1)}}{(n+1)!\ },
\end{align}
and the series mathematically converges for any $t$. However, the truncation error at order $N \geq n^*$ takes the form
\begin{align}
    \varepsilon_N(t) = \Big\|\sum_{n=N+1}^\infty \Theta_n(t)\Big\| \lesssim A\,\frac{N_s^{s_0-3/2}}{e^{N_s}} \frac{e^{4N_sn_dt}[4N_sn_dt]^{N+2}}{(N+2)!},
\end{align}
where we have used the Lagrange form of the reminder of the Taylor expansion of $e^x = \sum_k x^k/k!$. The convergence is thus slow and the requirement for staying within a target tolerance $\epsilon$,  $\varepsilon_N(t)\le\epsilon$ gives
\begin{equation}
\label{eq:Ntruncation_exp}
    A N_s^{s_0-3/2}e^{-N_s} e^{4N_sn_dt}[4N_sn_dt]^{N^*+2} \leq \epsilon (N^*+2)!,
\end{equation}
which requires $N^* \gtrsim 4N_sn_dt \sim N_s t$.
\\\indent
\paragraph{Truncation bounds for long time gradient.}
From the error bound calculated above, we can estimate the error bound on the series truncation when the evaluation of the dynamics is divided in $n_I$ equal subintervals $[t_j, t_j + \Delta t]$, with $t_0 = 0$ and $t_{n_I} = t_0 + n_I \Delta t = t$. From \cref{eq:unitarycompositederivative} we saw that the gradient can be expressed as
\begin{align}\label{eq:gradsplittingform}
    G \equiv \nabla_{\bm{\alpha}} U(t) = U(t) \sum_j U^\dagger(t_j) \left[ \sum_n \Theta_n(t_j, t_{j-1}) \right] U(t_j).
\end{align}
We note that using \cref{eq:gradsplittingform} uses the same iterative series as obtained in \cref{eq:gradientseriesexpansion}. Thus the only change when calculating $\Theta_n(t_j, t_{j-1})$ happens at the limits of integration for $\beta_{\textbf{m},m}^{(n)}(t_{j-1},t_j)$, yielding the bound 
\begin{align}
\label{beta_bound_t_t0_SM}
\big| \beta_{\bm{m}, m}^{(n)}(t_{j-1},t_j) \big| &{}\leq \frac{1}{n!} \int_{t_{j-1}}^{t_j} dt' \ (t_j-t')^{n} \left| \frac{\partial u_m(t')}{\partial\alpha_k} \right|
\leq \frac{1}{n!} \int_0^{\Delta t} d\tau \ \tau^{n} \left| \frac{\partial u_m(t_j-\tau)}{\partial\alpha_k}\right| \nonumber\\
&{}\leq \max_{t'\in[t_{j-1},t_j]}\left|\frac{\partial u_m(t')}{\partial\alpha_k}\right|\frac{1}{n!} \int_0^{\Delta t} \tau^n d\tau = c_j\,\frac{(\Delta t)^{n+1}}{(n+1)!},
\end{align}
with $\Delta t = t_j - t_{j-1}$ and $c_j = \max_{t'\in[t_{j-1},t_j]}|\partial u_m(t')/\partial\alpha_k|$.
Lets denote $\Theta(t_j, t_{j-1}) = \sum_{n=0}^\infty \Theta_n(t_j, t_{j-1})$ as the exact series, and $\Tilde{\Theta}(t_j, t_{j-1}) = \sum_{n=0}^{N_j} \Theta_n(t_j, t_{j-1})$ as the truncated series by considering $N_j$ terms. Using the considerations above, the errors related to each interval $j$ will differ from each other only by the maximum value of $|\partial u_m(t')/\partial\alpha_k|$ in $[t_{j-1}, t_j]$ and the chosen truncation $N_j$:
\begin{align}
    \norm{\Theta(t_j, t_{j-1}) - \Tilde{\Theta}_{N_j}(t_j, t_{j-1})} < \epsilon_j = c_j\,\tilde\epsilon_{N_j}(\Delta t).
\end{align}
When choosing the same truncation $N_j = N$ for all the intervals, the error bound on the gradient $\Tilde{G}$ obtained using the truncated series $\{\Tilde{\Theta}(t_j, t_{j-1})\}$ is given by
\begin{align}
    &\norm{G - \Tilde{G}} = \norm{U(t) \sum_j U^\dagger(t_j) \left[ \Theta(t_j, t_{j-1}) - \Tilde{\Theta}_{N_j}(t_j, t_{j-1}) \right] U(t_j)} \nonumber\\
    &\leq \norm{U(t)} \sum_j \norm{U^\dagger(t_j) \left[ \Theta(t_j, t_{j-1}) - \Tilde{\Theta}_{N_j}(t_j, t_{j-1}) \right] U(t_j)} \leq \sum_j \epsilon_j \nonumber\\
    &\leq n_I \tilde\epsilon_{N}(\Delta t) \max_{t'\in[0,t]}\left|\frac{\partial u_m(t')}{\partial\alpha_k}\right| \sim n_I \frac{(r\Delta t)^{N+2}}{1 - r\Delta t},
\end{align}
with $r\sim 4n_d$. The second line uses $\norm{U(t)} = 1$ and unitary evolution conserves the norm, i.e., $\norm{U^\dagger O U} = \norm{O}$.
Since $\Delta t = t/n_I$, the result above highlights that if on one side we can make the series convergent by choosing a number of intervals $n_I$ such that $rt/n_I < 1$, on the other the error on the total truncation can be also reduced in the same way, for it scales as $1/n_I^{N+1}$.

\section{Locality and Lieb-Robinson bound}\label{sec:localityandliebrobinson}
As the spread of the support of the evolution of a local operator is limited under a geometrically local Hamiltonian, a bound on the growth of the support of the series \cref{eq:gradientseriesexpansion} can also be computed.
The growth of support of an operator under a local Hamiltonian is, more formally, captured by the Lieb-Robinson bound. 
This finite speed of operator growth justifies the restriction of the dynamics simulation to a bounded subspace. 

For instance, let us consider a QOC problem of the form \cref{eq:hqoc} with the additional assumptions that the time-independent Hamiltonian $H_0$ is two-local, while the control terms $H_d(t) = \sum_m u_m(t) H_m$ act only on a driven region $\Lambda_0$. The system is initialized in a product state $\rho(0) = \ket{\psi(0)}\bra{\psi(0)} = \rho^{\Lambda_0}(0) \otimes \ket{0}\bra{0}^{\bar{\Lambda}_0}$, with $\bar{\Lambda}_0$ the complement of $\Lambda_0$. In the interaction picture, the propagator takes the form $U(t) = U_0(t) U_I(t)$ with $U_0(t) = e^{-i H_0 t}$ the free lattice evolution and $U_I(t)$ the interaction-picture propagator generated by $H_I(t) = U_0^\dagger(t) H_d(t) U_0(t)$. Thus, the support of $U_I(t)$ grows in time under the local interactions of $H_0$, within a region $\Lambda_{\mathrm{LR}}(t)$ which can be bounded by the Lieb-Robinson bound, i.e. $\text{supp}\{U_I(t)\} = \Lambda_{\mathrm{LR}}(t)$. After a time $t$, the Hilbert space can be truncated to a region 
$\Lambda = \Lambda_{\mathrm{LR}}(t) + l$, with $l$ a buffer region that ensures the validity of the approximation. 
Defining the target state as $\sigma = \ket{\psi_f}\bra{\psi_f}$, for unitary dynamics the state fidelity takes the form
\begin{equation}
\label{eq:fidelity_dms}
    \mathcal{F}(t) = \mathrm{tr}\left\{ U_0(t) U_I(t)\rho(0)U_I^\dagger(t) U_0^\dagger(t) \sigma \right\} = \mathrm{tr}\left\{\rho_I(t) \sigma_0(t)\right\} ,
\end{equation}
with $\sigma_0(t) = U_0^\dagger(t) \sigma U_0(t)$. We choose $\sigma_0(t)$ so that it factorizes as $\sigma_0(t)= \sigma_t^\Lambda \otimes \ket{0}\bra{0}^{\bar{\Lambda}}$, with $\bar{\Lambda}$ the complement of $\Lambda$. Given the local support of $U_I(t)$, we can write $U_I(t) \simeq U_I^\Lambda(t)\otimes \mathbb{I}^{\bar{\Lambda}}$, and thus the evaluated fidelity can be restricted to the bounded subspace $\Lambda$,
\begin{equation}
\label{eq:fidelityrestricted}
    \mathcal{F}(t) = \mathrm{tr}\left\{ U_I^\Lambda(t)\rho^{\Lambda}(0) U_I^{\Lambda,\dagger}(t) \, \sigma_t \otimes \ket{0}\bra{0}^{\bar{\Lambda}} \right\} =  \mathcal{F}_{\Lambda}(t) .
\end{equation}

In what follows, we discuss possible truncation of the series based on the Lieb-Robinson light cone, and briefly discuss the Magnus expansion for local Hamiltonians to locally approximate $U(t)$ in \cref{eq:gradientseriesexpansion}.

\subsection{Lieb-Robinson bound}
\label{app:lbbound}
The Lieb-Robinson bound, originally derived in Ref.~\cite{lieb_finite_1972}, is a measure of the speed of correlation growth in locally interacting quantum systems. Here we show how the Lieb-Robinson bound can be used to determine possible numerical truncation strategies of $\partial_\alpha U(t)$. Specifically, let us consider the right-hand side of \cref{eq:thetaseriestoint}. Using basic properties of the time evolution operator we can write it as
\begin{align}
\label{eq:inttminustau}
     -i U(t)  \int_{0}^t d \tau \frac{\partial u_m (\tau)}{\partial \alpha_k} H_m^{(\mathrm{H})}(\tau) U^{\dagger}(t) = 
      -i \int_{0}^t d \tau \frac{\partial u_m (\tau)}{\partial \alpha_k}  U(t, \tau) H_m U^{\dagger}(t, \tau),
\end{align}
Let $O_B$ be a generic operator with support on a subset $B$ of the qubits. Our goal is to bound the norm of the commutator of the integral in \cref{eq:inttminustau} with $O_B$. Using the triangle inequality we immediately obtain:
\begin{align}
\label{eq:inttminustaubound}
    \left \lVert \left [ -i \int_{0}^t d \tau \frac{\partial u_m (\tau)}{\partial \alpha_k} U(t, \tau) H_m U^{\dagger}(t, \tau), O_B \right] 
     \right \rVert \le 
      \int_{0}^t d \tau \left \lvert\frac{\partial u_m (\tau)}{\partial \alpha_k} \right \rvert  \left \rVert \left [ U(t, \tau) H_mU^{\dagger}(t, \tau), O_B \right] \right \rVert.
\end{align}
Thus, our goal is to obtain a bound for $\left \rVert \left [ U(t, \tau) H_mU^{\dagger}(t, \tau), O_B \right] \right \rVert$, which is exactly the main object studied in the Lieb-Robinson bound. Several formulations of the Lieb-Robinson bound are available in the literature~\cite{hastings2004, bravyi2006, hastingskoma2006,  hastings2010, poulin2010, barthel2012, kliesch_lieb-robinson_2014, chen2023}. 
Here we use the result from Barthel, Kliesch \cite{barthel2012}, which is stated for time-dependent generators and for evolutions $U(t,\tau)$ over an arbitrary window $[\tau, t]$. 

Let us consider a many body system where $V$ denotes the set of all subsystems, which in our case can be taken as qubits. We consider time-dependent Hamiltonians on this system of the following form
\begin{equation}
\label{eq:hgeneral}
    H(t) = \sum_{A} H_A(t),
\end{equation}
where the sum is over subsets $A$ of the subsystems and $H_A(t)$ denote terms that act non-trivially only on subsystems in $A$. We assume that the system is provided with a distance $d(a, b) \ge 0$ for $a, b \in V$. For instance, if the system is associated with a graph the distance can be the shortest path between the $a$ and $b$. We note however that this does not imply directly that \cref{eq:hgeneral} is defined on a graph as in \cref{eq:graphHamiltonian} . A distance between subsystems can be used to define the distance $d(a, B)$ between a subsystem $a$ and a subset $B$, the distance $d(A, B)$ between two subsets $A$ and $B$, and the diameter $\mathrm{diam}(A)$ of a subset $A$:
\begin{subequations}
\begin{align}
    d(a, B) &= \mathrm{min}_{b \in B} d(a, b), \\
    d(A, B) &= \mathrm{min}_{a \in A, b \in B} d(a, b), \\
    \mathrm{diam}(A) &= \mathrm{max}_{a, a' \in A} d(a, a').
\end{align}
\end{subequations}

Following Barthel, Kliesch \cite{barthel2012}:
Let $O_A$ and $O_B$ be two operators with support on subsets $A$ and $B$, respectively. For unitary dynamics, the Lieb-Robinson bound is a bound on the norm of the commutator $\left[U^\dagger(t,t_0)O_AU(t,t_0), O_B\right]$. Following the results of Ref.~\cite{barthel2012}, for a Hamiltonian system with couplings described by a graph $G=(V,E)$ with maximum degree (local connections) $\Delta$, we have that
\begin{align}
\label{eq:LRboundtimedependenthamiltonian}
    \norm{\left[U^\dagger(t,t_0)O_AU(t,t_0), O_B\right]} &\leq 2 m_{A,B} \norm{O_A} \norm{O_B} e^{v_{\mathrm{LB}} \abs{\,t-t_0} - d(A, B)},
\end{align}
where $m_{A,B} = \min{\big\{|A|/\Delta, |B|/\Delta\big\}}$, with $|W|:= |\{Z\subset W | H_Z(t)\neq 0 \}|$ and we have defined the Lieb-Robinson velocity $v_{\mathrm{LB}}$ as
\begin{equation}
\label{eq:vlbgraph}
    v_{\mathrm{LB}} = e\,\Delta\sup_{t'\in[t_0,t],A}\norm{H_A(t')}.
\end{equation}

We can now apply the bound in \cref{eq:LRboundtimedependenthamiltonian} to the integrand on the right-hand side of \cref{eq:inttminustaubound}. Let $A_m$ denote the support of $H_m(t)$. We obtain
\begin{align}
\label{eq:lbboundseries}
\begin{split}
    &\left \lVert \left [ -i \int_{0}^t d \tau \frac{\partial u_m (\tau)}{\partial \alpha_k} U(t, \tau) H_m U^{\dagger}(t, \tau), O_B \right]
     \right \rVert \leq  2\, m_{A_m, B}\norm{H_m} \norm{O_B} \, e^{v_{\mathrm{LB}}t - \, d(A_m, B)} \int_0^t d\tau \, \left| \frac{\partial u_m(\tau)}{\partial\alpha_k} \right|  e^{ - v_{\mathrm{LB}}\tau}  \\
    &{} \leq 
    2 m_{A_m, B} \|H_m\| \|O_B\|  e^{v_{\mathrm{LB}}t - \, d(A_m, B)} \left[ \int_0^t d\tau \, \left|\frac{\partial u_m(\tau)}{\partial\alpha_k} \right|^2 \right]^{1/2} \left[ \int_0^t d\tau \, e^{ -2 v_{\mathrm{LB}}\tau} \right]^{1/2} \\
    &{} = 
    2 m_{A_m, B} \|H_m\| \|O_B\| e^{v_{\mathrm{LB}}t - \, d(A_m, B)} \sqrt{t} \left [ \mathrm{max}_{\tau\in[0,t]} \left| \frac{\partial u_m(\tau)}{\partial\alpha_k} \right|^2 \right ]^{1/2} \left[
     \frac{1}{2 v_{\mathrm{LB}}} \left(1 - e^{- 2 v_{\mathrm{LB}} t} \right) \right]^{1/2} \\
    &{} = 
    \frac{ m_{A_m, B} }{\sqrt{v_{\mathrm{LB}} }}\|H_m\| \|O_B\| \left [ \mathrm{max}_{\tau\in[0,t]} \left| \frac{\partial u_m(\tau)}{\partial\alpha_k} \right|^2 \right ]^{1/2} \sqrt{2t
     \left(e^{2 v_{\mathrm{LB}} t} - 1 \right) } e^{ - \, d(A_m, B)} \\
    &{} \leq 
    \frac{ m_{A_m, B} }{\sqrt{v_{\mathrm{LB}} }}\|H_m\| \|O_B\| \left [ \mathrm{max}_{\tau\in[0,t]} \left| \frac{\partial u_m(\tau)}{\partial\alpha_k} \right|^2 \right ]^{1/2} \sqrt{2t}\,
    e^{ - [ d(A_m, B) - v_{\mathrm{LB}} t]},
\end{split}
\end{align}
which for short times  $v_{\mathrm{LB}} t \ll 1$ simplifies to :
\begin{equation}
\label{eq:lbsmalltime}
    \left \lVert \left [ -i \int_{0}^t d \tau \frac{\partial u_m (\tau)}{\partial \alpha_k} U(t, \tau) H_m U^{\dagger}(t, \tau), O_B \right]
     \right \rVert \lesssim  2\, m_{A_m, B} \|H_m\| \|O_B\| \left [ \mathrm{max}_{\tau\in[0,t]} \left| \frac{\partial u_m(\tau)}{\partial\alpha_k} \right|^2 \right ]^{1/2} t\, e^{ - \, d(A_m, B)}.
\end{equation}

Thus, for small total time, the left-hand side in \cref{eq:lbsmalltime}, which is the quantity of interest in our series expansion, has an exponentially decreasing support on subsets $B$ with increasing distance $d(A_m, B)$.
\cref{eq:lbsmalltime} suggests that the operators considered in the series should have support on sites $b \in V$ that are within distance $d(A_m, b) \le \mathcal{O}(\ln (t))$.

We note that the Lieb-Robinson bound in this form is a weak bound, and tighter bounds could be formulated \cite{mcdonough_lieb-robinson_2025, else_improved_2020, wang_tightening_2020}. We emphasize here that in lieu of the bounds derived here, we apply heuristic strategies motivated by the bounds derived in this, and the previous section, for numerical truncation of the series that are tighter.
Although tighter bounds could indicate the amount of resources needed for the computation of the gradients, derivation of tighter Lieb-Robinson bounds is outside the scope of this work.

\subsection{Magnus expansion and Light cone}\label{suppsection:MagnusExpansion}
In the previous section, we motivated the truncation of the series $\Theta_n$ (in \cref{eq:gradientseriesexpansion})  to a smaller subspace, determined by the Lieb-Robinson bound. Computing the gradient of the cost function, as given by \cref{eq:parderfidelity}, requires one to additionally compute the unitary propagator, and correspondingly the fidelity. The unitary propagator $U(t)$ generally spans the entire Hilbert space.  A possible way for computing it is by means of the Magnus expansion~\cite{blanes2009}. Ref.~\cite{sharma2024hamiltoniansimulationinteractionpicture} demonstrates that under a geometrically local Hamiltonian, the Magnus operator can be approximated by a local Magnus operator defined on a smaller Hilbert space, i.e.,
\begin{align}
    \norm{e^{\Omega(t)} - e^{\Omega_\text{loc}(t)}} \leq \norm{\Omega - \Omega_\text{loc}} \sim \bigO{t e^{-d/2}},
\end{align}
with $\Omega_\text{loc}(t)$ defined by dropping terms with a distance greater than $d$. Combining this with the bound expressed in \cref{eq:lbboundseries} could potentially point towards resources needed for the control of geometrically local quantum systems. We leave this for future work, and instead use the $4^\text{th}$- and $6^\text{th}$-order approximation to the Magnus expansion with Gauss-Legendre quadratures  for computing the unitary propagator as detailed in Ref.~\cite{blanes2009}.

For the $4^\text{th}$-order Magnus expansion, one needs to evaluate $[H_0, H_m]$ and $[H_{m'}, H_m]$. And for the $6^\text{th}$-order, along with the previous two terms, two more commutators need to be evaluated using the minimum commutator formula \cite{blanes2009}. Using locality of the drive operators, the $4^\text{th}$ and $6^\text{th}$-order expansion are reduced to one and three commutators, respectively. Finally, we emphasize that some of the commutators computed for the series could be reused for the Magnus expansion, and fewer commutators and coefficients need to be computed in each iteration.

\section{Connection with other QOC methods and extensions}\label{sec:connectionwithotherqoc}
In this section, by starting from \cref{eq:parderintegralform}, we demonstrate a straightforward derivation of the exact gradient in GRAPE \cite{defouquieres2011}. This helps to build intuition for the series derived in \cref{eq:gradientseriesexpansion} as a generalization of the exact GRAPE gradient to any continuous pulse parameterization. Finally, in \cref{Supp section: formal solution for nth order derivative}, we extend our previous gradient derivation, and show that the formal solution for any $n^\text{th}$-order derivative can be derived from first principles.

\subsection{Exact derivatives in GRAPE}\label{Supp: connection with GRAPE}
We assume that the time-dependent Hamiltonian is given by a sequence of $N_I$ time-independent, yet parameterized, Hamiltonians:
\begin{equation}
\label{eq:hpwc}
    H(t) = H_j(\bm{\alpha}_j), \quad t \in [t_{j -1}, t_j)
\end{equation}
with $j = 1, \dots, N_I$, $t_0 = 0, t_{N_I}=t$. We also fix a constant time interval $\Delta t = t_j - t_{j-1}, \, \forall j$. Each vector of parameters $\bm{\alpha}_j$ may have $N_j$ parameters which we denote as $\alpha_{j k}$. With this assumption the derivative of the propagator can be written as
\begin{align}
    \frac{\partial U(t)}{\partial \alpha_{jk}} = U(t, t_j) \frac{\partial U(t_j, t_{j-1})}{\partial \alpha_{jk} }U(t_{j-1}, t_0).
\end{align}
The derivative on the right-hand side can be evaluated by means of \cref{eq:parderintegralform} 
\begin{equation}
    \frac{\partial U(t_j, t_{j-1})}{\partial \alpha_{jk}} = -i \int_{t_{j-1}}^{t_{j}} d \tau \, U(t_j, \tau) \frac{\partial H(\tau)}{\partial \alpha_{jk}}  U(\tau, t_{j-1}),
\end{equation}
and using \cref{eq:hpwc} we obtain
\begin{align}
    \frac{\partial U(t_j, t_{j-1})}{\partial \alpha_{jk}} = -i\Delta t \ U_j \int_0^1  ds \  U_j^{-s} \frac{\partial H_j}{\partial \alpha_{jk}} U_j^s,
\end{align}
with $U_j = e^{-i H_j \Delta t}$. Note that we left the parameterization arbitrary and thus $\dajk H_j$ does not necessarily commute with $H_j$. Using the BCH expansion \cite{hall_book} the integrand can be written as a series 
\begin{align}
    e^X (\dajk H_j) e^{-X} = \sum_{n=0}^\infty \frac{\mathcal{C}_X^n (\dajk H_j)}{n!},
\end{align}
where $\mathcal{C}_X (\dajk H_j) = [X, \dajk H_j]$ is the superoperator corresponding to the commutator with $X$, and $\mathcal{C}_X^n (\dajk H_j) = [X, [X, [... [X, \dajk H_j]]]]$ represents $n$ nested commutators, with $X = i H_j s \Delta t$.
Performing this substitution we recover the exact gradient as 
\begin{align}\label{appendix eq: GRAPE exact gradient series}
\begin{split}
    \frac{\partial U(t_j, t_{j-1})}{\partial \alpha_{jk}} &= -i \Delta t \ U_j \int_0^1 ds \left\{ \sum_{n=0}^\infty  \frac{\left( i s \Delta t \right)^n }{n!} \mathcal{C}_{H_j}^n(\dajk H) \right\}\\
    & = -i \Delta t \ U_j \sum_{n=0}^\infty  \frac{\left( i \Delta t \right)^{n}}{(n+1)!} \mathcal{C}_{H_j}^n(\dajk H) \\
    & = -i \Delta t \ U_j \ \gamma_j(\dajk H),
\end{split}
\end{align}
where 
\begin{equation}
\gamma_j= \frac{e^{\mathcal{C}_{i H_j \Delta t}} - 1}{\mathcal{C}_{i H_j \Delta t}}
\end{equation}
represents the series for the superoperator $\mathcal{C}_{i H_j \Delta t} $ and matches the derivative in \cite{defouquieres2011}.

\subsection{Formal solution for \texorpdfstring{$n^\text{th}$}{nth}-order derivative}\label{Supp section: formal solution for nth order derivative}
The coupled differential equation for the $n^\text{th}$-order derivative can be derived by repeatedly differentiating the Schr\"odinger equation with respect to some parameter $\alpha$. For the second derivative, 
\begin{align}
\begin{split}
     \dt U &= -i H U \\ 
    \partial_\alpha(\dt U) &= -i (\partial_\alpha H) U - i H (\partial_\alpha U) \\
     \partial_\alpha^2(\dt U) &= -i \big( (\partial_\alpha^2 H)U + 2(\partial_\alpha H)(\partial_\alpha U) + H (\partial_\alpha^2 U) \big),
\end{split}
\end{align}
with $\partial_\alpha^2 U \equiv \frac{\partial^2 U}{\partial \alpha^2}$.
Rearranging the last equation, and switching the order of derivatives on the left-hand side,
\begin{align}
   \dt(\partial_\alpha^2 U) + i H (\partial_\alpha^2 U) = -i \left[ (\partial_\alpha^2 H)U + 2(\partial_\alpha H)(\partial_\alpha U)  \right].
\end{align}

Noting that this equation has a similar structure as \cref{eq:goatgrad}, we convert this to a homogeneous differential equation by left multiplying both sides by $U^\dagger(t)$. In this way, using
\begin{align}
\begin{split}
    \partialder{}{t}(U^\dagger \partial_\alpha^2 U) &= U^\dagger \dt(\partial_\alpha^2 U) + (\dt U^\dagger) (\partial_\alpha^2 U)   =U^\dagger \dt(\partial_\alpha^2 U) + i U^\dagger H (\partial_\alpha^2 U)
\end{split}
\end{align}
to obtain
\begin{align}
    \partialder{}{t}(U^\dagger \partial_\alpha^2 U) = -i U^\dagger(\partial_\alpha^2 H)U - 2i U^\dagger (\partial_\alpha H)(\partial_\alpha U).
\end{align}
Integrating both sides and left multiplying with $U(t)$ on both sides, we get
\begin{align}
\begin{split}
    (\partial_\alpha^2 U)(t) = & -i U(t) \int_0^t d \tau \, U^\dagger(\tau) \qty(\partial_\alpha^2 H(\tau)) U(\tau) -2i U(t)\int_0^t  \tau \, U^\dagger(\tau) \qty(\partial_\alpha H(\tau)) (\partial_\alpha U)(\tau).
\end{split}
\end{align}
Expanding $\partial_\alpha U(\tau)$ in the above equation using \cref{eq:gradintegralform}, we obtain the formal solution for the second-order derivative
\begin{align}
    \partial_\alpha^2 U(t) =  -i \int_0^t d \tau \, U(t, \tau) \qty(\partial_\alpha^2 H(\tau)) U(\tau) + 2 \qty(-i)^2 \int_0^t  d \tau' \int_0^{\tau'} d \tau'' \, U(t, \tau') \qty(\partial_\alpha H(\tau')) U(\tau', \tau'') \qty(\partial_\alpha H(\tau'')) U(\tau'').
\end{align}
Higher-order derivatives could be evaluated in a similar manner. We notice that the formal solution involves all compositions of $n$; counting the number of such combinations, the formal solution for the $n^\text{th}$-order derivative can be written as \cref{eq:formalsolutionfornthderivative}. This can be proved straightforwardly by induction.

\section{Numerical implementation}\label{sec:numericalimplementation}
In this section, we discuss the numerical gradient computation from the formal solution, \cref{eq:parderintegralform}, and the series expansion, \cref{eq:gradientseriesexpansion}. We outline a sparse matrix implementation using Pauli strings to store the commutator tree, and present the optimization results of the two model systems discussed in the main text.

\subsection{Code Implementation}

In this work, we presented the formal solution for the gradient of the unitary propagator, \cref{eq:parderintegralform}, and the series expansion of the formal solution, \cref{eq:gradientseriesexpansion}. Moreover, we highlighted that the gradient of the cost function for state-transfer problems can be computed by forward- and reverse-propagation of the initial and the target states, \cref{eq:parderfidelity}. In this section, we briefly discuss the numerical implementation of these three methods. 

Computing the gradient of the propagator using \cref{eq:parderintegralform} involves propagating the gradient of the Hamiltonian in the Heisenberg picture, as shown in Algorithm~\ref{Algo2:Gateoptimisationusingintegral}. 
In this case, all the operators corresponding to the subsystems whose parameters need to be optimized can be propagated in parallel. This algorithm requires storing $(m+1)$ dense matrices in memory, where $m$ is the number of operators being propagated.
Furthermore, gradient of all the parameters associated with a particular operator can also be computed simultaneously. This results in a highly parallelizable method, although one has to consider the time-step $\Delta t$ small enough to accurately estimate the integral.

For state-preparation tasks, computing the gradient using  \cref{eq:parderfidelity} involves computing the forward propagation of the initial state and the backward propagation of the target state, similar to GRAPE. A pseudo-code for this method is shown in Algorithm~\ref{Algo3:Statetransferclosedsystem}.
This method also enables parallel computation of the gradient for all parameters associated with a particular operator. As one only needs to propagate the states, ODE based methods can also be used in this case. Finally, similar to the above case, the time-step $\Delta t$ needs to be small enough to evaluate the integral accurately.

\begin{algorithm}[t]\label{Algo2:Gateoptimisationusingintegral}
    \caption{Gate optimization by \cref{eq:parderintegralform}}
    $t \gets t_0$\;
    $U(t) \gets \mathbb{I}$\;
    \While{$t < t_{\mathrm{final}}$}{
    \tcp{Using Magnus expansion}
    $U(t) \gets e^{\Omega(t + \Delta t, t)} U(t)$\;
    $\nabla_{\bm{\alpha}} \gets \nabla_{\bm{\alpha}} + U(t)^\dagger \qty(\da H(t)) U(t)$\;
    $t \gets t + \Delta t$\;
    }
    $\nabla_{\bm{\alpha}} \gets (-i \Delta t) U(t_\mathrm{final}) \, \nabla_{\bm{\alpha}}$\;
    \Return $\nabla_{\bm{\alpha}}$
\end{algorithm}

To compute the gradient using the series expansion, \cref{eq:gradientseriesexpansion}, the optimization process involves computing and storing a commutator tree, which in each iteration is multiplied with an analogous tree for the coefficients $\beta^{(n)}_{\bm{m}, m}(t)$ to obtain the series, as highlighted in the main text. 
For optimization of processes involving long final time, \cref{Main text eq: long time gradients} is used to split the series into shorter time lengths. As the series for these shorter time lengths are independent, these can be computed in parallel to speed up the computation. Moreover, for numerical computation of the series, we normalize the Hamiltonian such that $\norm{H} < 1$, and scale the final time accordingly; a standard requirement in some commonly used QOC methods \cite{ndong2010achebychevpropagator, reich2012monotonicallyconvergent}.
Finally, the propagator is computed using the Magnus expansion, as mentioned in \cref{suppsection:MagnusExpansion}, by reusing terms from the same commutator tree computed above.

Although the size of the commutator-tree can grow exponentially fast with increasing order, we use truncation strategies based on locality and some heuristic metrics, similar to sparse Pauli dynamics \cite{rudolph2025paulipropagationcomputationalframework, begusic2025realtimeoperatorevolution}, to reduce the number of commutators needed to obtain a good approximation.  
We prune leaves of the commutator tree based on \emph{graph topology}, \emph{operator weight}, and the \emph{branching order}. 
Truncation based on the graph topology takes into account the distance between the qubits, and the spread of the operator $\da H$. For short times, the spread of $\da H$ is localized, as shown in \cref{app:lbbound}.
Here, we compute the shortest distance ($d$) between two qubits (for instance at site $m$ and $l$) from an input connectivity graph. Based on this distance, we determine if the support of the drive operator at site $m$ reaches site $l$.
As the drive operators $H_m$ are assumed to be one- or two-local, the support of $H_m$ can only increase by commutation with other two-local terms in the Hamiltonian. 
Thus, using a quantity that we call the \emph{radius of the drift (drive) Hamiltonian} ($r$), we say that the support of $H_m$ reaches a site $l$ after $n$ commutation with the drift (non-local drive) if the distance $d < nr$, else the commutator with $H_l$ would be zero. Hence, we can prune all the terms with $d > nr$. 

On the other hand, operator weight and branching order based truncation are heuristic methods.
Truncation based on operator weight takes into account the number of subsystems onto which the terms in the commutator list act non-trivially, and ignores terms if the number of nontrivial operators is more than the operator weight.
This is based on the observation that $\beta$s corresponding to terms with high operator weight are small, as they involve a product of multiple drive pulses. For truncation based on branching order, we ignore commutators with the drive operator if the order of the term
is higher than the branching order. This is based on the observation that, for terms with large order of commutators, the highest contribution is from the commutator
with the drift Hamiltonian.

Moreover, we can safely ignore commutators with operators from subsystems that are not driven, as the coefficient for such terms would be zero.
This can be seen from \cref{eq:betaintegralform}, where setting $u_{m^{(n)}}(\tau) = 0$ makes $\beta_{\bm{m}, m}^{(n)}(t) = 0$.
Thus, in the commutator-tree, we only need to consider the subsystems with non-zero drive during that time period. This further restricts the growth of the commutator-tree. While the truncation strategies mentioned here typically reduce the size of the commutator-tree, they may not be applicable to all kinds of systems. Hence, our algorithm also supports a user-supplied list of relevant commutator indices, which in certain cases can drastically reduce the number of terms needed to be computed.

A pseudo-code for the algorithm is shown in Algorithm~\ref{Algo1:Gateoptimisationusinggradientseries}.
The implementation of all three algorithms is written in Python, based on the open-source package ParaQeet \cite{paraqeet}. We use JAX \cite{jax2018github} to JIT-compile parts of the algorithms for better performance.

\begin{algorithm}[t]\label{Algo3:Statetransferclosedsystem}
    \caption{State-transfer problem by \cref{eq:parderfidelity}}
    $t \gets t_0$\;
    $\ket{\psi(t)} \gets \ket{\psi_0}$\;
    $\bra{\psi_f(t)} \gets \bra{\psi_f}$\;
    $\da \gets 0$\;
    
    \While{$t < t_{\mathrm{final}}$} {\tcp{Forward propagation}
    $ \ket{\psi(t)} \gets e^{-i H(t) dt} \ket{\psi(t)} $\;
    $t \gets t + \Delta t$\;
    }
    
    \While{$t > t_0$}{\tcp{Backward propagation}
    $ \ket{\psi(t)} \gets e^{i H(t) dt} \ket{\psi(t)} $\;
    $ \bra{\psi_f(t)} \gets \bra{\psi_f(t)} e^{-i H(t) dt} $\;
    $\da \gets \da -i \, \Delta t \, \mel{\psi_f(t)}{\da H(t)}{\psi(t)} $\;
    $t \gets t - \Delta t$\;
    }

    \tcp{Evaluate state-overlap}
    $\Phi \gets \bra{\psi_f}\ket{\psi(t_{\mathrm{final}})}$\;
    $\da \gets \da \cdot \Phi^* + \da^* \cdot \Phi $\;
    
    \Return $\da$
\end{algorithm}

\subsection{Details of the simulations}\label{sec:latticesimulation}

As the Hamiltonian, and the commutators between the terms in the Hamiltonian, are sparse compared to the unitary operator, memory requirements for optimization of locally interacting qubits can further be reduced by storing them as sparse matrices. Using the standard binary representation of Pauli operators (see Ref.~\cite{rudolph2025paulipropagationcomputationalframework} for reference), we encode Pauli strings as a tuple of two $n$-bit binary numbers. The sparse matrices can then be represented as Python dictionaries
\begin{align}
    A = \{ \bm{r} : \chi_A(\bm{r}) \}, \ \forall \bm{r} \ \text{s.t.} \ \chi_A(\bm{r}) \neq 0 ,
\end{align}
where $\bm{r} \equiv (\bm{x}, \bm{y})$ represents a Pauli string $P(\bm{r})$ encoded in terms of two $n$-bit binary numbers $\bm{x}$ and $\bm{y}$. The value $\chi_A(\bm{r}) \equiv \Tr{A \, P(\bm{r})}$ represents the characteristic function of operator $A$. With the operators represented in the Pauli basis, basic matrix operations like matrix multiplication and commutators can then be performed by bit-wise binary operations.
Finally, these sparse matrices are used to store the commutator tree, and only the unitary propagator is computed as a dense matrix. For simplicity, we use our own python implementation for Pauli strings; for more performant alternatives refer to Ref. \cite{pauli_engine}. 

\begin{figure}
    \centering
    \includegraphics{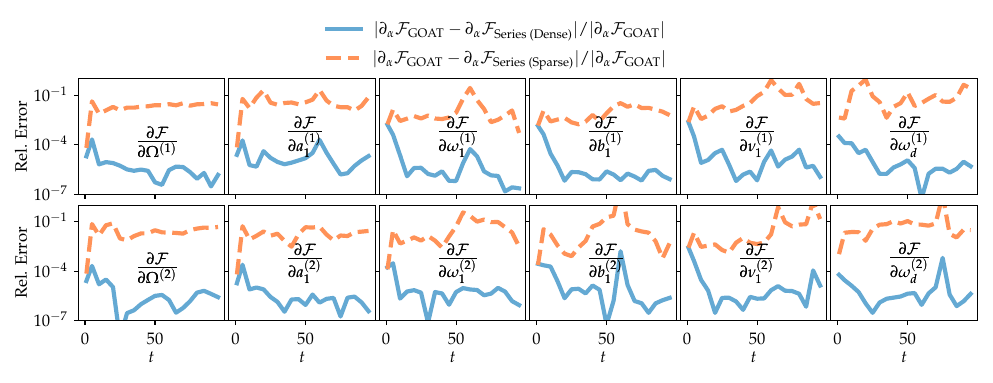}
    \caption{Relative error in gradient of the fidelity with respect to pulse parameters --- amplitude $(\Omega)$, Fourier coefficients $(a^{(i)}_j, b^{(i)}_j)$, Fourier frequency $(\omega^{(i)}_j, \nu^{(i)}_j)$, drive frequency $(\omega_d^{(i)})$ --- for two qubits with a dCRAB envelope along with sinusoidal oscillations, corresponding to \cref{Main text fig: comparision of gradients}. Here the gradient computed using the dense and the sparse implementation of the series is compared with the gradient computed using GOAT. It can be seen that in most of the cases the derivatives match up to a high accuracy even with an operator weight of 3.}
    \label{fig:gradientcomaprisonreldiff}
\end{figure}

As discussed in the main text,  we validate our method by considering a 1D chain of qubits with drive on the first and second qubits, with the aim to prepare a Bell state in the presence of spectator qubits, i.e., $\ket{\psi_f}=\left( \ket{01} + \ket{10} \right)/\sqrt{2} \otimes \ket{0}^{\otimes (n-2)}$, by starting from state $\ket{0}^{\otimes n}$. \cref{Main text fig: comparision of gradients} shows a comparison with the GOAT method~\cite{machnes2018} for the gradient of a two-qubit system with a CRAB envelope, including sinusoidal oscillations, as pulse ansatz, given by 
\begin{align}\label{eq:crabpulseansatz}
\mathcal{E}^{(k)}(t) =  \Omega^{(k)} \left(\sum_{j = 1}^{N_c} \left( a_j^{(k)} \cos(\omega_j^{(k)} t + \phi_j^{(k)}) + i b_j^{(k)} \cos(\nu_j^{(k)} t + \xi_j^{(k)}) \right) \right) \cos(\omega_d^{(k)} t) ,
\end{align}
where $\mathcal{E}^{(k)}(t)$ represents the drive on the $k^\text{th}$ qubit, with $N_c$ being the number of Fourier components in the CRAB envelope; and applied as a local $\mathrm{X}$-drive. Here the optimizable parameters are the amplitude $(\Omega^{(k)})$, Fourier coefficients $(a^{(k)}_j , b^{(k)}_j)$, Fourier frequency $(\omega^{(k)}_j, \nu^{(k)}_j)$, and drive frequency $(\omega^{(k)}_d)$. 
We compute the gradient for some chosen parameters, as shown in \cref{Main text fig: comparision of gradients}(a), of both the drives and compare the gradient obtained from the series expansion, implemented using dense and sparse matrices as discussed in the next section, with the ones from GOAT and show a good correspondence. The series is computed with an operator weight of 3, branching order of 7, $\Delta t$ of 10, final time 100, and commutators with drift are considered till order 30.
We can see that all the derivatives match up to a relative error of $10^{-4}$ for the dense implementation and $10^{-1}$ for the sparse implementation, as demonstrated in \cref{fig:gradientcomaprisonreldiff}.
Moreover, in \cref{Main text fig: comparision of gradients}(b) it can be seen that for large system sizes the sparse implementation requires less memory at the expense of higher wall time, as expected. 
The deviation in the gradient values for the case of sparse implementation is attributed to discarding Pauli strings with small coefficients from the sparse representation. Finally, we note that all three methods compared in \cref{Main text fig: comparision of gradients}(b) were optimized for speed at the expense of memory. Nevertheless, we expect similar trends would hold when using memory-efficient implementations.

\subsection{Optimization}
To further verify the accuracy of the gradient, in the following section we consider a model system of a 1D chain of qubits and perform optimization to construct a Greenberger–Horne–Zeilinger (GHZ) state in the presence of spectator qubits. A similar problem for a ladder geometry is discussed in the main text with results shown in \cref{fig:modelsystem2dladder}.

\begin{figure}
    \centering
    \includegraphics{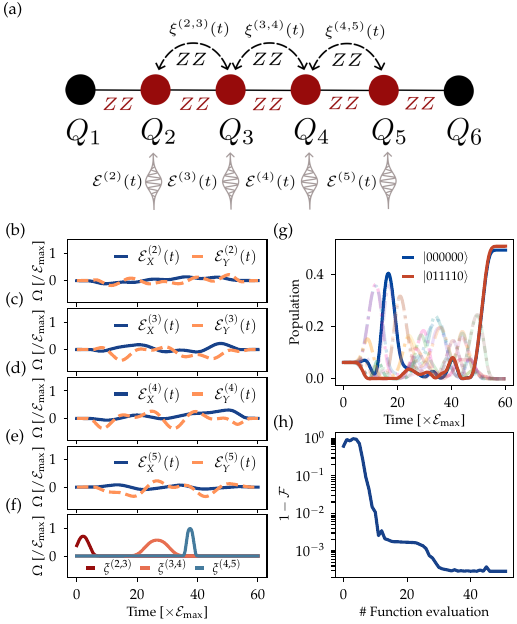}
      \caption[GHZ state-preparation on a 1D qubit chain]{\textbf{Model system:} GHZ state-preparation on a 1D qubit chain. (a) schematics of the model of a 1D chain of 6 qubits, with the aim to construct a GHZ state on $Q_2$, $Q_3$, $Q_4$, and $Q_5$ denoted in red by using the tunable $\mathrm{ZZ}$ interactions $(\xi^{(i,j)}(t))$ and the single qubit drives $(\mathcal{E}^{(j)}_{X/Y}(t))$. Qubits $Q_1$ and $Q_6$, shown in black, act as spectator qubits with stray $\mathrm{ZZ}$ interactions between all qubits. (b-e) The optimized pulse shapes for the local $\mathrm{X}, \mathrm{Y}$ drives and (f) the optimized pulse shapes for the tunable $\mathrm{ZZ}$ interactions between the qubits, with a total of 260 optimizable parameters. (g) Corresponding dynamics, highlighting only the computational states relevant for the target GHZ state. (h) Infidelity for each function evaluation during the L-BFGS-B optimization, starting from a pulse optimized without stray $\mathrm{ZZ}$ interactions.}
    \label{fig:modelsystem1dchain}
\end{figure}

We consider a chain of 6 qubits with the aim to construct a 4-qubit GHZ state on the four center qubits, while the other two act as spectators, as demonstrated in \cref{fig:modelsystem1dchain}(a). In this model, we assume that each qubit can be driven by a local drive for single qubit operations, and a tunable $\mathrm{ZZ}$ interaction between the nearest-neighbors for two qubit operations. Furthermore, as a source of imperfection in the model, we also consider always-on stray $\mathrm{ZZ}$ interactions between the nearest-neighbors. The Hamiltonian of this model can be written in a frame rotating at the frequency of the qubits as
\begin{align}
    H(t) = \sum_{i = 2}^{5} \left(\mathcal{E}_X^{(i)}(t) X^{(i)} + \mathcal{E}_Y^{(i)}(t) Y^{(i)}\right)  + \sum_{i = 2}^{4} \xi^{(i, i+1)}(t) Z^{(i)} Z^{(i + 1)} + \sum_{i = 1}^{5} J^{(i, i+1)} Z^{(i)} Z^{(i+1)}.
\end{align}
where $\mathcal{E}_X^{(i)}(t)$ and $\mathcal{E}_Y^{(i)}(t)$ are resonant local $\mathrm{X}$- and $\mathrm{Y}$-drive on the qubit, $\xi^{(i, i+1)}(t)$ the tunable $\mathrm{ZZ}$ drive, and $J^{(i, i+1)}$ the stray $\mathrm{ZZ}$ coupling strength. Here the $J^{(i, i+1)}$ values are chosen randomly to represent imperfection in the model.
For the pulse ansatz, we assume that the single qubit drives are characterized with CRAB pulses, given by \cref{eq:crabpulseansatz}, with $N_c = 10$, and the $\mathrm{ZZ}$ drives with flat-top Gaussian pulses, resulting in a total of $260$ optimizable parameters. A gate optimization was performed for the following circuit 
\begin{center}
\begin{quantikz}
    \lstick{$\ket{+}$} & \ctrl{1}  & \qw       & \qw       & \qw      & \qw  & \qw  & \qw\\
    \lstick{$\ket{+}$} & \ctrl{-1} & \gate{H}  & \ctrl{1}  & \qw      & \qw  & \qw  & \qw\\
    \lstick{$\ket{+}$} & \qw       & \qw       & \ctrl{-1} & \gate{H} & \ctrl{1} & \qw & \qw \\
    \lstick{$\ket{+}$} & \qw       & \qw       & \qw       & \qw      & \ctrl{-1} & \gate{H} & \qw \\
\end{quantikz}
\end{center}
to obtain a GHZ for the initial state $\ket{0++++0}$. The optimized pulses and the dynamics are shown in \cref{fig:modelsystem1dchain}, with the stray coupling values being $J^{(1,2)} = 0.29, J^{(2,3)} = 0.13, J^{(3,4)} = 0.011, J^{(4,5)} = 0.02, J^{(5,6)} = 0.2$.
A gradient based L-BFGS-B optimization was performed with the gradient computed using \cref{eq:gradientseriesexpansion} along with  \cref{Main text eq: long time gradients}, with order 5, operator weight 3, and a $\Delta t$ of 3.0. Since each qubit contains 3 drives, in this case we choose low-order for the series with a small $\Delta t$ to ensure an accurate gradient while simultaneously ensuring a fast evaluation. \cref{fig:modelsystem1dchain}(g) demonstrates the infidelity $(1 - \mathcal{F})$ with each function evaluation during the L-BFGS-B optimization, by starting with an optimized pulse shape without any stray $\mathrm{ZZ}$ interactions, and re-optimizing after adding the stray $\mathrm{ZZ}$ terms. The optimization was stopped manually at an infidelity of $0.00029$. It can be seen that the infidelity decreases quickly, indicating the accuracy of the gradient. 

A similar optimization is then also performed for a 2D ladder geometry of qubits, as shown in \cref{fig:modelsystem2dladder}. Here, we prepare a GHZ state on the four center qubits: $Q_2$, $Q_3$, $Q_6$, and $Q_7$. Qubits $Q_1$, $Q_4$, $Q_5$, and $Q_8$ act as spectator qubits. Here, the stray coupling values are $J^{(1,2)} = 0.29, J^{(2,3)} = 0.13, J^{(3,4)} = 0.01, J^{(4,5)} = 0.02, J^{(5,6)} = 0.20, J^{(6,7)} = 0.24, J^{(7,8)} = 0.26, J^{(1,8)} = 0.024, J^{(2,7)} = 0.24, J^{(3,6)} = 0.05$. The optimization was performed with order 3, operator weight 3, and a $\Delta t$ of 5.0, with a total of 264 parameters. The results of the optimization are shown in \cref{fig:modelsystem2dladder}, stopped manually at an infidelity of 0.0005.

\end{document}